\documentclass[lettersize,journal]{IEEEtran}
\pdfoutput=1
\usepackage{amsmath,amsfonts}
\usepackage{amsthm}
\usepackage{algorithmic}
\usepackage{algorithm}
\usepackage{array}
\usepackage{color}
\usepackage[caption=false,font=normalsize,labelfont=sf,textfont=sf]{subfig}
\usepackage{makecell}
\usepackage{textcomp}
\usepackage{tabularx}
\usepackage{stfloats}
\usepackage{url}
\usepackage{verbatim}
\usepackage{graphicx}
\usepackage{cite}
\usepackage{multirow}
\usepackage{multicol}
\usepackage{ragged2e}
\newtheorem{Lemma}{ \textit{Lemma}}
\newtheorem{ass}{Assumption}

\newtheorem{thm}{Theorem}

\usepackage{CJKutf8}
\begin{document}
\title{Performance Analysis of End-to-End LEO Satellite-Aided Shore-to-Ship Communications: A Stochastic Geometry Approach
%End to End LEO Satellite-Aided Shore-to-Ship Communication Networks: Modeling and Performance Analysis
}

\author{Xu Hu,~\IEEEmembership{Student Member,~IEEE}, Bin Lin,~\IEEEmembership{Senior Member,~IEEE}, Xiao Lu,~\IEEEmembership{Member,~IEEE}, Ping Wang, ~\IEEEmembership{Fellow,~IEEE}, Nan Cheng,~\IEEEmembership{Senior Member,~IEEE}, Zhisheng Yin,~\IEEEmembership{Member,~IEEE}, Weihua Zhuang,~\IEEEmembership{Fellow,~IEEE}
        % <-this % stops a space %Manuscript received May 24, 2022; accepted July 4, 2022. Date of publication May 9, 2022; date of current version July 11, 2022.
\thanks{
Manuscript received 12 July 2023; revised 3 December 2023 and 29 February 2024; accepted 24 March 2024. The work was supported by the National Natural Science Foundation of China (No. 62371085, No. 51939001), Fundamental Research Funds for the Central Universities (No. 3132023514), and partially supported by a Canada NSERC Discovery grant. An earlier version of this paper was presented in part at the 12th IEEE/CIC International Conference on Communications in China (ICCC 2023), Dalian, China, 2023 \cite{iccc}. The associate editor coordinating the review of this article and approving it for publication was Meng-Lin Ku. {\it{(Corresponding author: Bin Lin.)}}

X. Hu and B. Lin are with the Information Science and Technology College, Dalian Maritime University, Dalian 116026, China (e-mail: huxu@dlmu.edu.cn; binlin@dlmu.edu.cn). X. Lu is with the Research and Development, Ericsson, Ottawa, ON K2K 2V6, Canada (e-mail: lu9@ualberta.ca). P. Wang is with the Department of Electrical Engineering and Computer Science, York University, Toronto, ON M3J 1P3, Canada (e-mail: pingw@yorku.ca). N. Cheng is with the State Key Laboratory of ISN and the School of Telecommunications Engineering, Xidian University, Xi’an 710071, China (e-mail: dr.nan.cheng@ieee.org). Z. Yin is with the State Key Laboratory of ISN and the School of Cyber Engineering, Xidian University, Xi’an 710071, China (e-mail: zsyin@xidian.edu.cn). W. Zhuang is with the Department of Electrical and Computer Engineering, University of Waterloo, Waterloo, ON N2L 3G1, Canada (e-mail: wzhuang@uwaterloo.ca). 

This manuscript has been accepted by IEEE Transactions on Wireless Communications, DOI: 10.1109/TWC.2024.3384549.
}}

%\markboth{IEEE TRANSACTIONS ON WIRELESS COMMUNICATIONS, Vol.~X, No.~X, X~X}%{Shell \MakeLowercase{\textit{et al.}}: A Sample Article Using IEEEtran.cls for IEEE Journals}


\maketitle

\begin{abstract}
%Low Earth orbit (LEO) satellite networks offer strategic superiority in maritime communications, enabling signal transmissions from shore to ship through space-based links. The traditional performance modeling based on multiple circular orbits is challenging in large-scale LEO satellite constellations, requiring a tractable approach to evaluate the network performance. In this paper, we propose a theoretical framework for an LEO satellite-aided shore-to-ship communication network (LEO-SSCN), where LEO satellites are distributed as a binomial point process (BPP) on a specific sphere. The framework aims to obtain the end-to-end transmission performance by considering signal transmissions through either a marine link or a space link, which is subject to Rician or Shadowed Rician fading. Due to the indeterminate position of the serving satellite, we propose a distance approximation approach to accurately model the distance from the serving satellite to the destination ship. By incorporating a threshold-based communication scheme, we leverage stochastic geometry to derive analytical expressions of end-to-end transmission success probability and average transmission rate capacity. Extensive numerical results verify the accuracy of the analysis and demonstrate the effect of key parameters on the LEO-SSCN. \textcolor{blue}{Notably, with a 3 dB predefined threshold, the transmission success probability increases by 500${\rm{\% }}$ via incorporating the space link.}

Low Earth orbit (LEO) satellite networks have shown strategic superiority in maritime communications, assisting in establishing signal transmissions from shore to ship through space-based links. Traditional performance modeling based on multiple circular orbits is challenging to characterize large-scale LEO satellite constellations, thus requiring a tractable approach to accurately evaluate the network performance. In this paper, we propose a theoretical framework for an LEO satellite-aided shore-to-ship communication network (LEO-SSCN), where LEO satellites are distributed as a binomial point process (BPP) on a specific spherical surface. The framework aims to obtain the end-to-end transmission performance by considering signal transmissions through either a marine link or a space link subject to Rician or Shadowed Rician fading, respectively. Due to the indeterminate position of the serving satellite, accurately modeling the distance from the serving satellite to the destination ship becomes intractable. To address this issue, we propose a distance approximation approach. Then, by approximation and incorporating a threshold-based communication scheme, we leverage stochastic geometry to derive analytical expressions of end-to-end transmission success probability and average transmission rate capacity. Extensive numerical results verify the accuracy of the analysis and demonstrate the effect of key parameters on the performance of LEO-SSCN. Notably, with common parameter settings, after incorporating the space link, the transmission success probability increases by 886${\rm{\% }}$ with a 13 dB predefined signal-to-noise ratio (or signal-to-interference-plus-noise-ratio) threshold. This superior performance is attributed to the fact that the space link uses a wider bandwidth and greater power for signal transmission compared to the maritime link. It’s undeniable that the integration of the space link inevitably incurs additional expenses.  %with a 3 dB predefined threshold, the transmission success probability increases by 500${\rm{\% }}$ via incorporating the space link.
\end{abstract}

\begin{IEEEkeywords}
Low Earth orbit (LEO) satellites, shore-to-ship communications, stochastic geometry, binomial point process (BPP), end-to-end transmission success probability, average transmission rate capacity.
\end{IEEEkeywords}

\section{Introduction}
\IEEEPARstart{M}{aritime} activities have experienced significant growth in recent years, encompassing areas such as maritime transportation, tourism, search and rescue, resource exploration, and offshore aquaculture. As a result, there has been a tremendous surge in maritime data traffic from shore to ship \cite{refa1, refa2, refa3}. This necessitates the development of a high-speed, ultra-reliable, and low-latency shore-to-ship communication network to ensure seamless communication services \cite{refa2}. At present, data transmission services between shore and ships on the sea surface are primarily reliant on traditional maritime communication systems, such as the global maritime distress and safety system (GMDSS) \cite{GMDSS}, the navigation telex (NAVTEX) system \cite{Navtex}, the automatic identification system (AIS) \cite{VDES2}, the very high-frequency data exchange system (VDES) \cite{VDES2}, and narrow-band maritime satellite systems represented by the international maritime satellites (Inmarsats). However, these systems offer only low data rate messages or voice services \cite{refa1}. Furthermore, the limited radio spectrum resources, the harsh maritime propagation environment, and the restricted coverage present significant obstacles to shore-to-ship communications. These factors make it increasingly challenging to meet the growing demands of maritime data traffic.

Recently, low Earth orbit (LEO) satellite constellations have attracted considerable interest as a promising solution to enhance maritime communications by providing ubiquitous and global coverage. With hundreds to thousands of LEO satellites orbiting the Earth, these satellite constellations offer high speed, high throughput, and cost-effective communication capabilities \cite{na}. Many satellite constellation projects, such as Starlink, Kuiper, and OneWeb have already launched many LEO satellites \cite{refl3, refl4}, forming LEO satellite networks that can provide continuous broadband communication services. These networks can support a wide range of maritime applications, e.g., navigational position information broadcasting, marine image collection, real-time voice and video streaming for emergency rescue, and multimedia traffic for onboard passengers \cite{refa3}. The LEO satellite network introduces a new avenue and shows strategic superiority for signal transmissions from shore to ship, enabling a seamless transition between connectivities on the sea surface and over space. Thus, by collaborating with the existing shore-to-ship communication network, an LEO satellite-aided shore-to-ship communication network (LEO-SSCN) holds promise for enhancing maritime communication capabilities.

The characterization of network performance for the LEO-SSCN is of great importance because of the ultra-expensive cost of LEO satellite deployment and the limited and precious orbital resources. Thorough evaluation and prediction of network performance boundaries are crucial before proposing LEO satellite commercial plans to ensure their feasibility and effectiveness. However, to characterize the transmission performance of the LEO-SSCN, the traditional circular orbit grid model used in satellite networks, wherein satellites are positioned on a grid composed of multiple circular orbits, such as the Walker constellation, poses challenges in large-scale LEO satellite constellations \cite{d1}. Moreover, intricate system-level simulations have limitations such as long runtimes, complex operations, and high resource requirements, making them suitable only for small-scale satellite constellations. To address these challenges, a tractable theoretical approach is indispensable as a complementary element to obtain insights into the LEO-SSCN design guidance.

Stochastic geometry is a powerful mathematical tool that has been widely utilized in the analysis of cellular networks and can provide analytical results for crucial performance metrics \cite{SG, zhuang, wei}. Due to the advantage of enabling tractable modeling for various types of wireless networks and analyzing their properties, stochastic geometry holds promise for the LEO-SSCN. Accordingly, in this paper, we aim to provide an analytical framework to characterize the network performance of the LEO-SSCN by using stochastic geometry and illustrate the impact of key parameters on the end-to-end signal transmission performance.

\subsection{Related Works}
In recent years, there has been a growing interest in utilizing stochastic geometry to analyze the performance of LEO satellite networks. Existing studies have focused on both uplink network performance from terrestrial users to LEO satellites \cite{up1, up2, up3, up4} and downlink network performance from LEO satellites to terrestrial users \cite{d1, d2, d3, dd1, dd2, d4, dd3, d5, d6, d7}. Studies on the uplink network performance included analytical approaches for the uplink service coverage probability (CP) in a satellite-terrestrial network to obtain the relationship between the number of satellites and base station density by modeling ground devices or users as a Poisson point process (PPP) \cite{up1}\cite{up2}. Additionally, tractable formulas were derived in \cite{up3} to evaluate the service outage probability (OP) and expected normalized throughput from ground devices to satellites. In \cite{up4}, B. Al Homssi \emph{et al.} provided a framework for optimizing the uplink coverage for satellite constellations by either individually or jointly tuning the altitude and the beamwidth of the LEO satellites.

%\cite{ref2, ref3, ref4}
%various aspects of the LEO satellite networks, including the derivation of essential analytic theorems to simplify the performance analysis model \cite{bas1, bas2, bas3, bas4}, 
%For essential analytic theorem derivation, researchers have modeled the fixed number of satellites as being independently and uniformly distributed on different concentric spheres. This has allowed them to derive exact analytical expressions for the cumulative distribution functions (CDFs) of the nearest neighbor and the contact distance \cite{bas1}. Subsequently, the analytical expression of the CDF of the conditional contact angle is derived in \cite{bas2} by establishing connections with Dome Angle. Both the derived expressions can be used in studying the coverage probability of the LEO-aided communication networks and the routing among LEO satellites and other applications. Due to the short period each satellite dwells within the ground user's field of view, frequent handovers between satellites are necessary, leading to increased signaling overhead and possible service interruptions when establishing connections with ground users. To this end, a tractable approach for characterizing satellites' pass duration and the CDF and probability density function (PDF) of the Doppler magnitude are obtained in \cite{bas3} and \cite{bas4}, respectively. The analytical expressions are essential for the design of caching schemes and provide a means of evaluating the performance of satellite networks.

The downlink performance analysis of the LEO satellite network has been extensively studied using stochastic geometry, which models the LEO satellite constellation by a certain distribution. For instance, J. Park \emph{et al.} in \cite{d1} used a random point process to enable a tractable analysis of the interference statistics and the expected service area CP. Additionally, Ruibo Wang \emph{et al.} confirmed the accuracy of modeling the LEO satellite constellation as a binomial point process (BPP) by establishing a relationship between the BPP and the Fibonacci lattice \cite{d2}. Based on this, several researchers investigated the performance of the massive LEO satellite constellation to capture crucial performance metrics \cite{d3, d4, dd1, dd2, dd3}. In \cite{d3}, the CP and average data rate for a given number of satellites modeled as BPP on a sphere at a given altitude were analyzed. However, in actual satellite constellations, the satellite density varies with latitude on the sphere, leading to the adoption of a uniform distribution of LEO satellites on the orbital shell that takes into account both the parameters of the latitude and the propagation angle \cite{dd1}. To more accurately model the actual distribution of LEO satellites along with different latitudes, a non-homogeneous PPP with a proper intensity was selected in \cite{dd2} and \cite{d4}. Since signal transmissions from the nearest satellite may be subject to severe shadowing due to the blockage by nearby obstacles surrounding ground users, the shortest distance scheme was replaced by the best server scheme for the highest signal-to-interference-plus-noise-ratio (SINR) \cite{dd3}. To obtain fine-grained analysis of the CP, the meta distribution of signal-to-interference-ratio (SIR) or SINR was employed to provide the probability that a certain percentage of links can reach a target threshold \cite{d5}. In LEO satellite commercial plans, satellites are envisioned to deploy at multi-spheres with different altitudes in the constellation. In \cite{d6}, the expressions of CP were derived for a setup where satellites were deployed at different altitudes, and the effect of altitudes and numbers of satellites as well as gateway density on the network performance was studied. Additionally, in \cite{d7}, the OP and symbol error rate over downlink channels were deduced when users were randomly located in single-beam and multi-beam areas. 

To model LEO satellite networks accurately, several works have focused on establishing analytic frameworks to describe signal transmission performance \cite{d8, d9, d10}. Ruolin Wang \emph{et al.} in \cite{d8} developed a novel expression for CP, accounting for satellite beam coverage and atmospheric attenuation, incorporating Rayleigh fading in the satellite-to-ground channel. In\cite{d9}, Seong Ho Chae \emph{et al.} obtained exact analytical expressions for CPs and ergodic rates under Nakagami-$m$ fading, using the PPP model to optimize beamwidth control and balance beam coverage with network interference. While the classic Rayleigh and Nakagami-$m$ models offer simple expressions for analysis, they are less suitable for accurately characterizing satellite-ground links compared to Shadowed Rician (SR) fading. Dong-Hyun Jung \emph{et al.} investigated LEO satellite communication systems by employing SR fading for satellite-ground links \cite{d10}. However, this study did not delve into the impact of interfering satellites on the overall network performance. Different from these works, to establish a suitable model for the LEO satellite networks, our analytical framework focuses on analyzing signal transmission performance by considering the space link subject to SR fading, incorporating the influence of interfering satellites on network performance. This significantly increases the difficulty in deriving expressions of network performance due to the introduction of additional variables related to interfering satellites and the use of SR fading with complex expressions.

%Moreover, \cite{d8, d9, d10} While Dong-Hyun Jung \emph{et al.} in \cite{d10} delve into the study of downlink LEO satellite communication systems under the Shadowed-Rician Fading, their primary emphasis lies in the analysis of distance distributions of the serving satellite, along with the derivation of precise and approximated expressions for outage probability. It's worth noting that, unlike our contributions, it does not concentrate on analyzing the influence of interfering satellites on the overall system performance.

%To improve the signal transmission performance of the LEO satellite network, several studies have introduced various relays and adopted the multi-link cooperative communication scheme. Jia Ye \emph{et al.} investigate a space-air-ground integrated network, where high-altitude platforms and terrestrial base stations act as intermediaries for signal relaying between ground users and satellites, and propose a theoretical framework for the outage performance and the asymptotic outage probability for simplistic scenarios \cite{end1}. The transmission performance of a novel space-air-ground-sea integrated network structure comprising four types of relay links under the practical channel models is studied, focusing on the coverage probability \cite{end2}. 

As mentioned above, despite significant progress made in analyzing the performance of LEO satellite networks using the tool of stochastic geometry, most studies focused on the uplink from terrestrial users to LEO satellites or the downlink from satellites to terrestrial users. The role of satellites as relays for the end-to-end signal transmission from the source node to the destination node needs further research in stochastic geometry analysis. Furthermore, considering the impact of interfering satellites and incorporating the complex expression of the SR fading model along with several additional variables, the task of deriving analytical expressions for the network performance becomes more challenging. Notwithstanding meeting the actual characteristics, the modeling and analysis of the end-to-end signal transmissions bring new challenges, as signals often undergo forwarding through one or more hop relays. The collective influence of various network nodes further amplifies the complexity of performance analysis \cite{hop}. Additionally, end-to-end signal transmissions usually involve multiple links, posing challenges in modeling the correlation among parameters across different links.

%To our best knowledge, there are limited studies on maritime and underwater communications, especially concerning their unique characteristics \cite{underwater1, underwater2}. For maritime communications, in [32], the authors employ the structure of SAGSIN to improve signal transmission performance for surface stations over a large-scale and far-reaching ocean surface. For underwater communications, Jiajie Xu \emph{et al.} in \cite{underwater1} propose a new multitier underwater acoustic communication network model. This model aims to analyze and optimize the coverage probability by employing a realistic model for the communication channel. The study reveals that a single-tier network excels when dealing with underwater devices confined to a limited depth range. Conversely, a multi-tier network performs more effectively in scenarios involving a large range of underwater device depths. Conversely, a multi-tier network proves more effective in scenarios involving a large range of underwater device depths. Furthermore, to improve the underwater communication capacity at a level of thousands of kilometers, the authors propose a three-hop underwater wireless acoustic communication structure to realize long-distance communication based on the actual sound fixing and ranging (SOFAR) channel model \cite{underwater2}. By specifying the densities of transmitters, relay stations, and receivers, this study culminates in deriving the final coverage probability between a transmitter and receiver through the three-hop link.

To the best of our knowledge, there is a scarcity of research on maritime and underwater communications \cite{end2, underwater1, underwater2}. For maritime communications, in \cite{end2}, the authors employed the structure of space-air-ground-sea integrated networks (SAGSIN) to improve signal transmission performance for surface stations over a large-scale and far-reaching ocean surface. For underwater communications, Jiajie Xu \emph{et al.} in \cite{underwater1} proposed a new multitier underwater acoustic communication network model, where the distribution of the surface station and underwater Internet of Things (UIoTs) devices was modeled as PPPs at different depths. By considering signal transmissions from the tagged surface station to the UIoT subject to the underwater acoustic path loss, the CP was derived for a typical UIoT device as a function of the densities of PPPs, the length of tiers’ tether length, and depth of UIoT devices. Furthermore, to improve the underwater communication capacity at a level of thousands of kilometers, the authors in \cite{underwater2} proposed a three-hop underwater wireless acoustic communication structure to realize long-distance communication based on the actual sound fixing and ranging (SOFAR) channel model. With the given densities of transmitters, relay stations, and receivers, the CPs of the three hops were analyzed, and the final CP from a transmitter to a receiver through the three-hop link was derived. Due to the unique characteristics, which differ from terrestrial communications, the maritime communication environment features fewer propagation obstacles. This limitation restricts the applicability of the Rayleigh channel model, commonly adopted in terrestrial communications. In maritime communications, signal transmissions are influenced by reflections, refractions, and diffractions on the sea surface caused by sea wave fluctuations. Attenuations induced by unique maritime environmental factors such as rainfall, fog, and sea waves have impacts on signal transmissions. The movements of ships exacerbate the rapid fluctuations in channel conditions, thereby complicating the characterization of multipath propagation.

\subsection{Contributions and Organization}
In this paper, we present an analytical framework for evaluating the performance of end-to-end signal transmissions from shore to ship in the LEO-SSCN. In particular, the network architecture of the LEO-SSCN consists of a network service center (NSC), a shore base station (BS), an Earth station (ES), a destination ship (DS), and an LEO satellite constellation where LEO satellites are modeled as a BPP uniformly distributed on a specific spherical surface. The NSC deployed on the land acts as the source node in the LEO-SSCN and is set at point (0, 0, ${r_{\rm{e}}}$), where ${{r_{\rm{e}}}}$ is the radius of the Earth. The DS on the sea surface acts as the destination node in the LEO-SSCN and is set at an arbitrary Euclidean distance away from the NSC. We consider the end-to-end signal transmission from the NSC to the DS through either a marine link or a space link, which is subject to Rician fading\cite{Rician-sea1} or SR fading\cite{d10, end2}, respectively. Since accurately modeling the distance from the serving satellite to the DS is challenging due to the indeterminate position of the serving satellite, we propose a tractable distance approximation approach. Then, we consider a threshold-based communication scheme, where the link selection relies on the relationship between the received signal-to-noise ratio (SNR) of the pilot signal in the marine link and the predefined threshold value. Therefore, by employing the distance approximation approach and incorporating the communication scheme, we leverage the tool of stochastic geometry to analyze the end-to-end transmission performance encompassing the end-to-end transmission success probability and average transmission rate capacity, and the performance improvement of the LEO-assisted transmission compared with the marine link. The main contributions of this study are summarized as follows. 

\begin{itemize}
	\item We propose a theoretical framework for the LEO-SSCN to analyze the end-to-end signal transmission performance, including the end-to-end transmission success probability and average transmission rate capacity for an arbitrarily located DS on the sea surface by utilizing stochastic geometry, incorporating the Laplace transform of the interfering satellites’ power. 
 
	\item We construct a geometry structure among the NSC, the serving satellite, and the DS and propose a tractable distance approximation approach to estimate the distance from the serving satellite to the DS by utilizing the Pythagorean theorem. This aims to address the problem that accurately modeling the distance from the serving satellite to the destination ship becomes intractable due to the indeterminate position of the serving satellite.

% Since the distance from the serving satellite to the destination ship is intractable, we construct a geometry sketch among the network service center, serving satellite and destination ship and propose a tractable distance approximation approach to estimate the distance from the serving satellite to the destination ship by utilizing the Pythagorean theorem.

\item We conduct extensive simulations to verify the accuracy of the derived expressions and quantitatively analyze the impact of key parameters on the end-to-end transmission performance of the LEO-SSCN, including constellation size, constellation altitude, and the distance from the BS to the DS. We also demonstrate the performance improvement of the LEO-assisted transmission compared to the marine link transmission and examine the impact of various parameters, such as the predefined SINR threshold and the distance from the BS to the DS, on this improvement. With common parameter settings, when we set the predefined SNR (or SINR) threshold at 13 dB, the transmission success probability experiences an increase of 886${\rm{\% }}$ by incorporating the space link. This superior performance is attributed to the fact that the space link uses a wider bandwidth and greater power for signal transmission compared to the maritime link. It’s undeniable that the integration of the space link inevitably incurs additional expenses. 
  \end{itemize}
  
\begin{figure}[t] 
 \center{\includegraphics[scale=0.55]{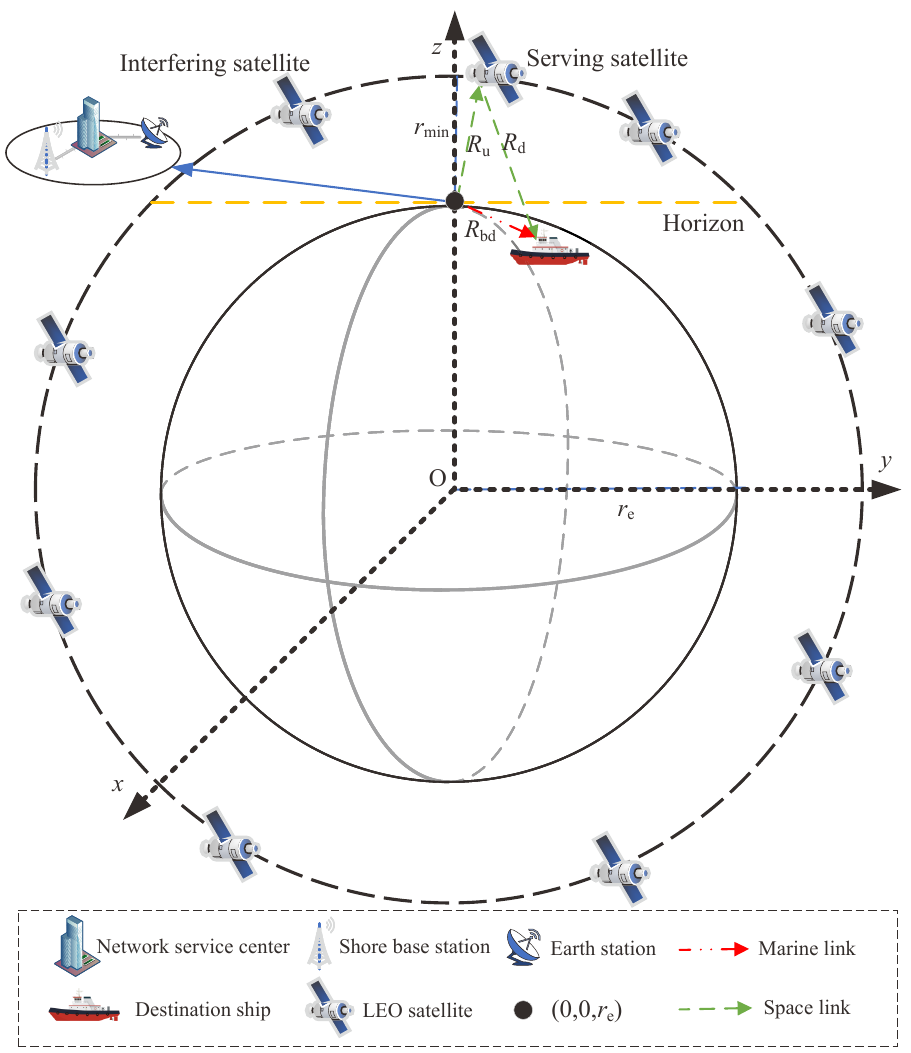}} %20031115-2
 \caption{\label{1} The geometric structure of the LEO-SSCN.} 
 	\label{fig:1}
\end{figure} 

The rest of this paper is organized as follows. Section \uppercase\expandafter{\romannumeral2} describes the geometric model, the communication scheme, and the channel model in the LEO-SSCN. Section \uppercase\expandafter{\romannumeral3} derives analytical expressions of the end-to-end transmission success probability and average transmission rate capacity for the LEO-SSCN. Extensive numerical results are provided in Section \uppercase\expandafter{\romannumeral4} to verify the derivations and investigate the effect of key parameters on the performance of the LEO-SSCN. Section \uppercase\expandafter{\romannumeral5} concludes this work.

\section{System Model}
In this section, we provide an overview of the LEO-SSCN by presenting its geometries and characteristics. We then present the communication scheme and the channel model employed for the end-to-end signal transmissions from shore to ship in the LEO-SSCN. Furthermore, we introduce the preliminary Lemmas for deriving analytical expressions of network performance.

\subsection{Geometric Model}
\begin{table*}[t]
	\caption{Tables of mathematical Acronyms}
	\label{parameter description}
	\begin{tabularx}{\textwidth}{c|l} %\multicolumn{1}{c}{}
		\hline
		\hline
		Notions     & \makecell[c]{Descriptions}          \\ \hline
		$N; K;{N_{\rm{I}}} $ & Constellation size; The number of orthogonal frequency channels in the space link; The number of ISs  \\ \hline
		${r_{\rm{e}}}$; ${r_{\min }}$; ${r_{\max }}$ & Earth radius; Constellation altitude; Maximum possible distance between the ES and a visible satellite    \\ \hline
		\multirow{2}{*}{$R_{\rm{bd}}$; $R_{\rm{u}}$; $R_{\rm{d}}$; $R_j$ } & Distance from the BS to the DS; Distance from the ES to the SS; Distance from  the SS to the DS; Distance from the ${j^{{\rm{th}}}}$ IS to\\& the DS         \\ \hline
	    ${p_{\rm{bd}}}$; ${p_{\rm{u}}}$; ${p_{\rm{d}}}$; ${p_{\rm{i}}}$ & Transmit power of the BS; Transmit power of the ES; Transmit power of the SS; Transmit power of each IS\\ \hline
        ${G_{\rm{bd}}}$; ${G_{\rm{u}}}$; ${G_{\rm{d}}}$; ${G_{\rm{i}}}$ & Antenna gain of the BS; Antenna gain of the ES; Antenna gain of the SS; Antenna gain of each IS \\ \hline
        \multirow{2}{*}{${H_{{\rm{bd}}}}$; ${H_{\rm{u}}}$; ${H_{\rm{d}}}$; ${H_j}$ } & Channel coefficient of the marine link; Channel coefficient of the uplink of the space link; Channel coefficient of the downlink of\\& the space link; Channel coefficient of the link from the ${j^{{\rm{th}}}}$ IS to the DS\\ \hline
		\multirow{2}{*}{$\sigma _{{\rm{bd}}}^2; \sigma _{{\rm{u}}}^2; \sigma _{{\rm{d}}}^2$} & Additive noise power in the marine link; Additive noise power in the first time slot of the space link; Additive noise power in the\\& second time slot of the space link \\ \hline
		${\alpha _{{\rm{bd}}}}; {\alpha}$ & Path loss exponent of the marine link; Path loss exponent of the space link\\ \hline	
        ${B_{\rm{bd}}}; {B_{\rm{esd}}}$ & Bandwidth of the channel in the marine link; Bandwidth of each channel in the space link  \\ \hline
		${P_{\rm{bd}}}; {P_{\rm{esd}}}; {P_{\rm{s}}}$ & Success probability in the marine link; Success probability in the space link; End-to-end transmission success probability \\ \hline
    	%${P_{{\rm{s|}}\Phi }}; {P_{{\rm{esd|}}\Phi }} $ & Conditional end-to-end success probability; Conditional success probability in the space link;  \\ \hline
		%${M_{\rm{s}}}^{ - 1}; M_{{\rm{esd}}}^{ - 1}$ & Mean local delay; Local delay in the space link;  \\ \hline
		${C_{\rm{bd}}}; {C_{\rm{esd}}}; {C_{\rm{s}}} $ & Capacity in the marine link; Capacity in the space link; Average transmission rate capacity   \\ \hline
		\hline
	\end{tabularx}
\end{table*} 
We consider LEO-SSCN that consists of an NSC, a BS, an ES, a DS, and an LEO satellite constellation. As shown in Fig. \ref{fig:1}, we establish a three-dimensional Cartesian coordinate system for the LEO-SSCN with the Earth's center as the origin, the vector pointing from the origin to the North Pole as the $z$-axis and the vector in the Earth's equatorial plane as the $y$-axis. The $x$-axis direction vector is perpendicular to both vectors $y$ and $z$. The NSC deployed on the land acts as the source node in the LEO-SSCN and is set at point $\left( {0,0,{r_{\rm{e}}}} \right)$. The DS, equipped with a medium frequency (MF) receiver and a satellite receiver, acts as the destination node in the LEO-SSCN and is set at an arbitrary Euclidean distance, henceforth referred to as distance, away from the NSC on the sea surface. The NSC is used for providing network services to the DS, i.g., navigational information, control commands, and various individual messages. It has two signal transmission interfaces, i.e., the BS and the ES, respectively. The BS is deployed on the coastline and can directly communicate with the DS. The ES, located near the coastline, is associated with the LEO satellite which passes. Both the BS and the ES are connected to the NSC via wires. The LEO satellite constellation contains $N$ LEO satellites that are uniformly distributed around the Earth at the same altitude ${r_{\min}}$, forming a BPP on a specific spherical surface with radius ${r_{\rm{a}}}{\rm{ = }}{r_{\rm{e}}} + {r_{\min }}$. Moreover, we consider each satellite equipped with a directional antenna that radiates its main lobe forming one beam towards the center of the Earth, and both the NSC and the DS are located within the same beam of an LEO satellite. The maximum possible distance between the ES and a satellite over its horizon, known as a visible distance, is represented by ${r_{\max }} = \sqrt {2{r_{\rm{e}}}{r_{\min }} + r_{\min }^2} $, and the corresponding LEO satellite is known as a visible satellite.

Particularly, there are two kinds of links in the LEO-SSCN for conducting the end-to-end signal transmission from shore to ship, i.e., the marine link and the space link, operating at MF and Ka bands, respectively. The marine link contains a single channel that facilitates signals directly transmitted from the NSC to the DS through the BS. The space link is established using the decode and forward (DF) protocol for signal transmissions from the NSC to the DS via an LEO satellite \cite{DF}. The spectrum of the space link is partitioned into $K$ ($K \le N$) orthogonal frequency channels, and each channel is randomly assigned to a subset of $N{\rm{/}}K$ satellites. Two time slots are assumed in the considered space link where the first time slot is associated with the uplink from the ES to an LEO satellite, and the downlink from the LEO satellite to DS is executed in the second time slot.

All signals are transmitted from the NSC to the DS for offering network services through either the BS in the marine link or the ES in the space link. The ES is associated with the nearest satellite, referred to as the serving satellite (SS). Other satellites in the same channel can cause interference in the vessel’s reception when they are above the horizon, denoted as interfering satellites (ISs) \cite{d3}. The distance between the BS and the DS is denoted by ${R_{{\rm{bd}}}}$, the distance between the ES and the SS by ${R_{{\rm{u}}}}$, the distance between the SS and the DS by ${R_{{\rm{d}}}}$, and we have ${R_{{\rm{u}}}} \gg {R_{{\rm{bd}}}}$ and ${R_{{\rm{d}}}} \gg {R_{{\rm{bd}}}}$. The number of ISs is denoted by ${N_{\rm{I}}}$ with ${N_{\rm{I}}} \le \frac{N}{K} - {\rm{1}}$. The distance from an IS to the DS is denoted by ${R_j}$, $j = 1,2,...,{N_{\rm{I}}}$. Other mathematical notions are summarized in Table \ref{parameter description}.

\subsection{Communication Scheme}
In the LEO-SSCN, we consider signal transmissions from the NSC to the DS governed by a threshold-based communication scheme. Specifically, the NSC should send the pilot signal \cite{pilot} to the DS via the marine link, and the DS should measure the SNR and report it to the NSC. After the pilot signal transmission, the NSC selects the marine link or the space link based on the SNR of the pilot signal at the DS. If the received SNR is larger than a predefined threshold value, the NSC transmits signals to the DS via the BS in the marine link. Otherwise, the NSC transmits signals via the ES to the SS in the first time slot of the space link, followed by the SS transmitting the signals to the DS in the second time slot of the space link.

\subsection{Channel Model}
For the marine link, the SNR at the receiver of the DS can be expressed as
\begin{equation} \label{snrbd}
{\rm{SN}}{{\rm{R}}_{{\rm{bd}}}} = \frac{{{p_{{\rm{bd}}}}{G_{{\rm{bd}}}}{{\left| {{H_{{\rm{bd}}}}} \right|}^2}R_{{\rm{bd}}}^{ - {\alpha _{{\rm{bd}}}}}}}{{\sigma _{{\rm{bd}}}^2}},
\end{equation}
\noindent where ${p_{{\rm{bd}}}}$ is the transmit power of the BS, ${G_{{\rm{bd}}}}$ is the antenna gain of the BS, $\sigma _{{\rm{bd}}}^{\rm{2}}$ is the received power of the additive noise in the marine link. In (\ref{snrbd}), $R_{{\rm{bd}}}^{ - {\alpha _{{\rm{bd}}}}}$ presents the power path loss model, where ${R_{{\rm{bd}}}}$ denotes the distance from the BS to the DS, ${\alpha _{{\rm{bd}}}}$ is the path loss exponent of the marine link. ${{H_{{\rm{bd}}}}}$ denotes the channel coefficient of the marine link. In maritime communications, the signal from the clear line-of-sight (LoS) path is the predominant signal with multipath components playing a weaker role in the overall signal strength. This scenario lends itself well suited to the Rician fading model, which is often used to describe one strong direct LoS component and many random weaker components \cite{digital1}. In contrast, the widely used Rayleigh fading model better suits non-LoS wireless communications, particularly in densely populated urban areas \cite{digital1}. Therefore, we employ the Rician fading model in the marine link, in which the cumulative distribution function (CDF) \cite{end1} can be expressed as 
\begin{equation}
{F_{{H_{{\rm{bd}}}}}}(x) = 1 - {Q_1}\left( {\frac{v }{\rho },\frac{x}{\rho}} \right),
\end{equation}
\noindent where ${Q_1}$ is the Marcum Q-function with parameters $v$ and $\rho$. That is, $v = \sqrt {\frac{{2{K_{\rm{Rician}}}}}{{2{K_{\rm{Rician}}} + 2}}}$, $\rho  = \sqrt {\frac{1}{{2{K_{\rm{Rician}}} + 2}}}$, in which ${{K_{\rm{Rician}}}}$ denotes the Rician K factor. Moreover, the probability density function (PDF) of the Rician fading model \cite{end1} is 
\begin{equation}\label{pdfrician}
    f_{{H_{{\rm{bd}}}}} \left( {x} \right) = \frac{x}{{{\rho ^2}}}\exp \left( {\frac{{ - \left( {{x^2} + {v^2}} \right)}}{{2{\rho ^2}}}} \right){I_0}\left( {\frac{{xv}}{{{\rho ^2}}}} \right),
\end{equation}
\noindent where ${I_0}$ is the modified Bessel function of the first kind.

For the space link, in the first time slot, the SNR at the receiver of the SS can be expressed as
\begin{equation}\label{snru}
    {\rm{SN}}{{\rm{R}}_{\rm{u}}} = \frac{{{p_{\rm{u}}}{G_{\rm{u}}}{{\left| {{H_{\rm{u}}}} \right|}^2}R_{\rm{u}}^{ - {\alpha}}}}{{\sigma _{\rm{u}}^2}},
\end{equation}
\noindent where ${p_{\rm{u}}}$ and ${G_{\rm{u}}}$ are the transmit power and the antenna gain of the ES, respectively, ${\sigma _{\rm{u}}^2}$ is the received power of the additive noise in the first time slot, $R_{\rm{u}}^{ - {\alpha}}$ presents the power path loss model, ${R_{\rm{u}}}$ is the distance from the ES to the SS, and ${\alpha}$ is the path loss exponent of the space link. When ${R_{\rm{u}}} > {r_{\max }}$, the SS is below the horizon of the ES and ${\rm{SN}}{{\rm{R}}_{\rm{u}}} = 0$. Note that ${{\left| {{H_{\rm{u}}}} \right|}^2}$ is the power gain of the uplink channel of the space link, employing the SR fading model. This choice is grounded in the unique challenges of satellite communications, where signal propagation suffers from various factors, including free space path loss, scattering, reflection, diffraction by buildings and other objects, shadowing, and multipath \cite{digital1}. To accurately model these characteristics, the SR fading model has been proposed and validated to fit very well with the published data \cite{simple}. Consequently, the SR fading model is utilized in the space link, in which the CDF can be expressed as \cite{d6}
\begin{equation}\label{cdfofsr}
\begin{aligned}
%\begin{array}{l}
{F_{|{H_{\rm{u}}}{|^2}}}(x) =& 1 - \mu \mathop \sum \limits_{n = 0}^{m - 1} \frac{{{{(1 - m)}_n}{{( - \delta )}^n}}}{{{{(n!)}^2}}}\mathop \sum \limits_{l = 0}^n \frac{{n!}}{{l!}}\\
& \times {x^l}{{\rm{e}}^{ - (\beta  - \delta )x}}{(\beta  - \delta )^{ - (n + 1 - l)}}.
%\end{array}.
\end{aligned}
\end{equation}
\noindent In (\ref{cdfofsr}), $\mu  = \frac{1}{{2b}}{\left( {\frac{{2bm}}{{2bm + \Omega }}} \right)^m}$, $\delta  = \frac{1}{{2b}}\left( {\frac{\Omega }{{2bm + \Omega }}} \right)$, $\beta  = \frac{1}{{2b}}$ with ${{b}}$, $m$ and ${\rm{\Omega }}$ being the half average power of the multi-path component, the Nakagami parameter and the average power of the LoS component, respectively, and ${\left( w \right)_n} = w\left( {w + 1} \right)...\left( {w + n - 1} \right)$ denotes the Pochhammer symbol.

For the second time slot of the space link, in addition to the impact of the additive noise, the signal transmissions from the SS to the DS are also influenced by ISs; therefore, the received SINR at the DS in the second time slot can be expressed as
\begin{equation}\label{sinrd}
{\rm{SIN}}{{\rm{R}}_{{\rm{d}}}}{\rm{ = }}\frac{{{p_{\rm{d}}}{G_{\rm{d}}}{{\left| {{H_{\rm{d}}}} \right|}^2}R_{\rm{d}}^{ - {\alpha}}}}{{\sigma _{\rm{d}}^2}+I_s},
\end{equation}
\noindent where ${p_{\rm{d}}}$ and ${G_{\rm{d}}}$ are the transmit power and the antenna gain of the SS, respectively, ${\sigma _{\rm{d}}^2}$ is the received power of the additive noise in the second time slot, $R_{{\rm{d}}}^{ - {\alpha}}$ represents the power path loss model, ${R_{\rm{d}}}$ is the distance from the SS to the DS, ${{{\left| {{H_{\rm{d}}}} \right|}^2}}$ is the power gain of the downlink channel of the space link, which follows the SR fading model, and $I_s$ is the cumulative interference power from all ISs. The interference is given by 
\begin{equation}
I_s \buildrel \Delta \over = \sum\limits_{j = 1}^{{N_{\rm{I}}}} {{p_{\rm{i}}}{G_{\rm{i}}}\left| {{H_j}} \right|^2}R_j^{ - {\alpha}}, 
\end{equation}
\noindent where $N_{\rm{I}}$ is a random variable denoting the number of interfering satellites, ${p_{\rm{i}}}$ and ${G_{\rm{i}}}$ are the transmit power and the antenna gain of each IS, respectively, ${\left| {{H_j}} \right|^2}$ is the power gain of the link from the ${j^{{\rm{th}}}}$ IS to the DS, which also follows the SR fading model, and ${R_j}$ denotes the distance from the ${j^{{\rm{th}}}}$ IS to the DS. By adopting the directional beamforming with fixed-beam antennas, we assume that the antenna gain of the serving satellite is concentrated in the main lobe, while the interfering satellites have their gains spread across side lobes. Let ${{\varphi}}$ represent the angle between the satellite and the DS. Consequently, the antenna gain of a satellite can be described as \cite{d10}
\begin{equation}
\begin{aligned}
G\left( \varphi  \right) = \left\{ {\begin{array}{*{20}{l}}
{{G_{\rm{d}}},\quad {\rm{ if}}\left| \varphi  \right| \le {\varphi _{{\rm{th}}}},}\\
{{G_{\rm{i}}},\quad {\rm{ otherwise}},}
\end{array}} \right.
\end{aligned}
\end{equation}
\noindent where ${\varphi _{{\rm{th}}}}$ is the threshold angle between the main and side lobes of the beam pattern.

\subsection{Preliminaries}
To derive analytical expressions of the network performance in the subsequent sections, we employ some preliminary Lemmas as follows and the proofs can be found in \cite{d3}.

\begin{Lemma}
\label{Lemma-1}
The CDF of distance ${R_{{\rm{u}}}}$ from the ES to the SS is given by
\begin{equation}\label{cdfrd}
\resizebox{0.99\hsize}{!}{$
\begin{aligned}
{F_{{R_{\rm{u}}}}}\left( {{r_{\rm{u}}}} \right) = \left\{ {\begin{array}{*{20}{l}}
{1 - {{\left( {1 - \frac{{r_{\rm{u}}^{\rm{2}} - r_{\min }^2}}{{4{r_{\rm{e}}}{r_{\rm{a}}}}}} \right)}^N},{r_{\min }} \le {r_{\rm{u}}} \le 2{r_{\rm{e}}} + {r_{\min }} ,}\\
{0,\qquad \qquad \qquad  \qquad {{\rm{otherwise}}}.}
\end{array}} \right.
\end{aligned}$}
\end{equation}
\end{Lemma}

\begin{Lemma}
\label{Lemma-2}
The PDF of distance ${R_{{\rm{u}}}}$ from the ES to the SS is given by
\begin{equation}
\resizebox{0.95\hsize}{!}{$
\begin{aligned}
{f_{{R_{\rm{u}}}}}\left( {{r_{\rm{u}}}} \right){\rm{ = }}\left\{ {\begin{array}{*{20}{l}}
{\frac{{{r_{\rm{u}}}N}}{{2{r_{\rm{e}}}{r_{\rm{a}}}}}{{\left( {1 - \frac{{r_{\rm{u}}^2 - r_{\min }^2}}{{4{r_{\rm{e}}}{r_{\rm{a}}}}}} \right)}^{N - 1}},{r_{\min }} \le {r_{\rm{u}}} \le 2{r_{\rm{e}}} + {r_{\min }}},\\
{0,\qquad \qquad \qquad \qquad \quad \rm{otherwise}}.
\end{array}} \right.
\end{aligned}$}
\end{equation}
\end{Lemma}

\begin{Lemma}
\label{Lemma-3}
Given ${R_{\rm{u}}} = {r_{\rm{u}}}$, the PDF of ${R_j}$ is given by
\begin{equation}
{f_{{R_j}\mid {R_{\rm{u}}}}}\left( {{r_j}\mid {r_{\rm{u}}}} \right) = \left\{ {\begin{array}{*{20}{l}}
{\frac{{2{r_j}}}{{4{r_{\rm{e}}}{r_{\rm{a}}} - r_{\rm{u}}^2 + r_{\min }^2}},{r_{\rm{u}}} < {r_j} < 2{r_{\rm{e}}} + {r_{\min }}},\\
{0, \qquad \qquad \quad \rm{otherwise}} .
\end{array}}\right.
\end{equation}
\end{Lemma}

\begin{Lemma}
\label{Lemma-4}
When the SS is at a distance ${r_{\rm{u}}}$ from the ES with ${r_{\rm{u}}} \ge {r_{\min }}$, ${N_{\rm{I}}}$ is a binomial random variable with parameters $\frac{N}{K} - 1$ and ${P_{\rm{I}}}$, that is ${N_{\rm{I}}} \sim {\rm{B}}\left( {\frac{N}{K} - 1,{P_{\rm{I}}}} \right)$, where $\frac{N}{K} - 1$ is the number of trials and ${P_{\rm{I}}}$ is the probability of a positive outcome in each trial, which can be expressed as 
 \begin{equation}
 {P_{\rm{I}}} = {{\left( {{r_{\min }} - \frac{{r_{\rm{u}}^2 - r_{\min }^2}}{{2{r_{\rm{e}}}}}} \right)} \mathord{\left/
 {\vphantom {{\left( {{r_{\min }} - \frac{{r_{\rm{u}}^2 - r_{\min }^2}}{{2{r_{\rm{e}}}}}} \right)} {\left( {2{r_a} - \frac{{r_{\rm{u}}^2 - r_{\min }^2}}{{2{r_{\rm{e}}}}}} \right)}}} \right.
 \kern-\nulldelimiterspace} {\left( {2{r_a} - \frac{{r_{\rm{u}}^2 - r_{\min }^2}}{{2{r_{\rm{e}}}}}} \right)}}. 
 \end{equation}
 \end{Lemma}

\section{Performance Analysis}
In this section, we focus on deriving analytical expressions of the end-to-end transmission success probability and average transmission rate capacity for an arbitrarily located DS on the sea surface by utilizing stochastic geometry, incorporating the Laplace transform of the interfering satellites’ power. Additionally, we present a tractable distance approximation approach to estimate the distance from the SS to the DS by the distance from the ES to the SS.

\subsection{The Proposed Distance Approximation Approach}\label{A0}
To derive analytic expressions for the end-to-end transmission success probability and average transmission rate capacity in the following sections, we first need to evaluate the CDF and the PDF of ${R_{\rm{d}}}$.

In the LEO-SSCN, due to the dense deployment of the LEO satellites, the SS can be approximately regarded as being directly above the NSC. On this basis, the distance from the ES to the SS and the distance from the SS to the DS are much larger than that from the BS to the DS, i.e., ${R_{\rm{u}}} \gg {R_{{\rm{bd}}}}$ and ${R_{\rm{d}}} \gg {R_{{\rm{bd}}}}$, thus the triangle among NSC, SS and DS in the constructed geometric structure can be approximately regarded as a right triangle. In this right triangle, ${R_{{\rm{d}}}}$ is regarded as the hypotenuse, and ${R_{\rm{u}}}$ and ${R_{{\rm{bd}}}}$ are regarded as the two legs.
 
\begin{figure}[t]
		\centerline{\includegraphics[scale=0.55]{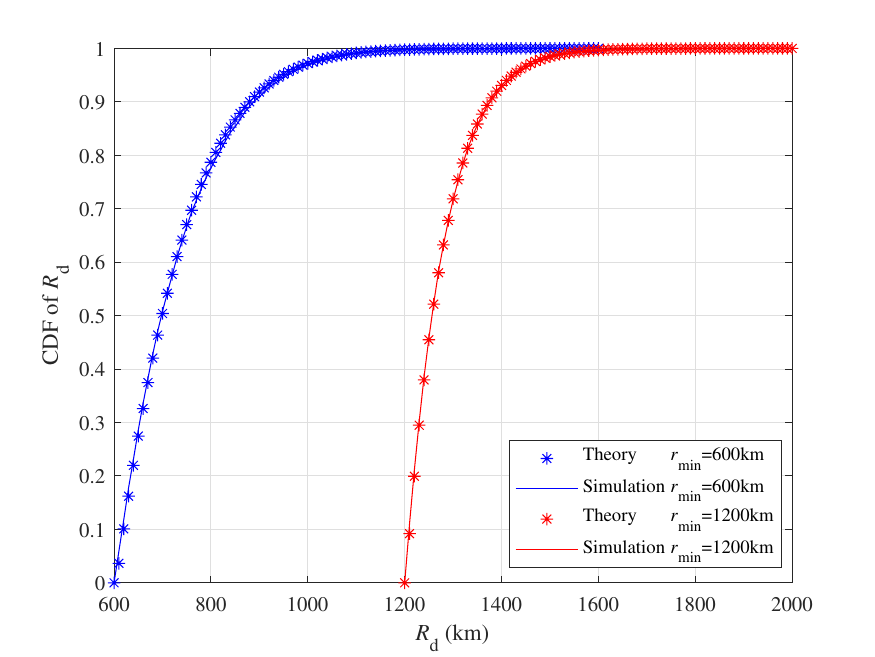}}%cdfrd1122
		\caption{The comparison of the approximated values and empirical results for CDFs of ${R_{\rm{d}}}$.}
		\label{CDF}
 \end{figure}

\begin{figure}[t]
		\centerline{\includegraphics[scale=0.55]{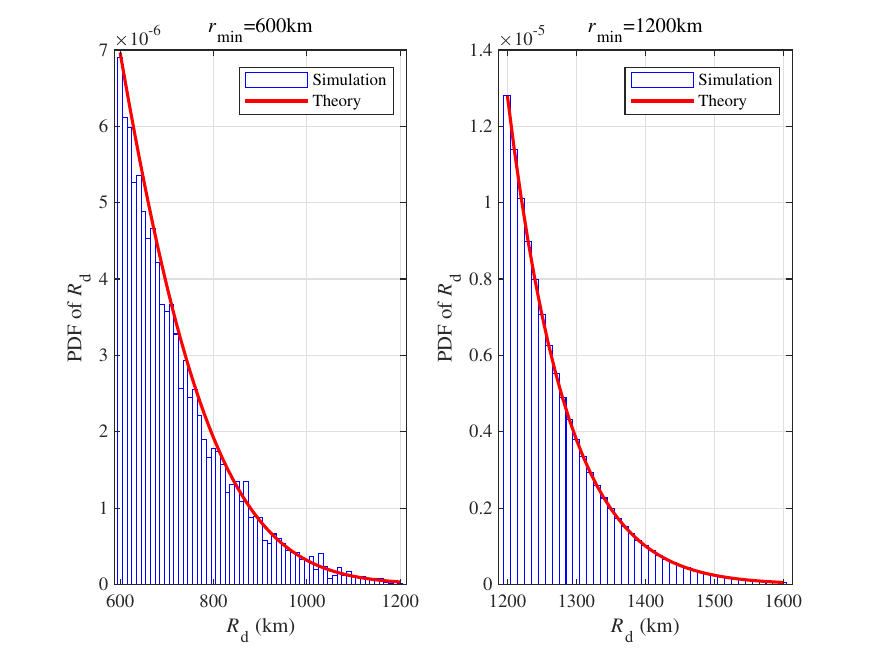}}%pdf600and1200-2
		\caption{The comparison of the approximated values and empirical results for PDFs of ${R_{\rm{d}}}$.}
		\label{PDF}
 \end{figure}

\begin{ass}
\label{pro-1}    
By utilizing the Pythagorean theorem, ${R_{\rm{d}}}$ can be approximated as 
\begin{equation}
{R_{\rm{d}}} \approx \sqrt { {R_{\rm{u}}^2 + R_{{\rm{bd}}}^2} } .
\end{equation}
\end{ass}
To validate the accuracy of the proposed approximation approach, we can obtain the CDF of ${R_{\rm{d}}}$ by substituting ${R_{\rm{u}}} \approx \sqrt { {R_{\rm{d}}^2 - R_{{\rm{bd}}}^2} } $ to ${R_{\rm{u}}}$ in Lemma 1, which can be expressed as %and (\ref{pdfrd}) in (\ref{cdfrd}), shown at the top of the next page.
\begin{equation}\label{cdfrd}
\resizebox{0.99\hsize}{!}{$
\begin{aligned}
{F_{{R_{\rm{d}}}}}\left( {{r_{\rm{d}}}} \right) = \left\{ {\begin{array}{*{20}{l}}
{1 - {{\left( {1 - \frac{{r_{\rm{d}}^{\rm{2}} - R_{{\rm{bd}}}^{\rm{2}} - r_{\min }^2}}{{4{r_{\rm{e}}}{r_{\rm{a}}}}}} \right)}^N},{r_{\min }} \le {r_{\rm{u}}} \le 2{r_{\rm{e}}} + {r_{\min }} ,}\\
{0,\qquad \qquad \qquad \qquad \qquad {{\rm{otherwise}}}.}
\end{array}} \right.
\end{aligned}$}
\end{equation}
\noindent Accordingly, the PDF can be obtained by taking the derivative of ${R_{\rm{d}}}$ in (\ref{cdfrd}), which can be expressed as 
\begin{equation}\label{pdfrd}
\resizebox{0.99\hsize}{!}{$
\begin{aligned}
{f_{{R_{\rm{d}}}}}\left( {{r_{\rm{d}}}} \right) = \left\{ {\begin{array}{*{20}{l}}
{\frac{{{r_{\rm{d}}}N}}{{2{r_{\rm{e}}}{r_{\rm{a}}}}}{{\left( {1 - \frac{{r_{\rm{d}}^{\rm{2}} - R_{{\rm{bd}}}^2 - r_{\min }^2}}{{4{r_{\rm{e}}}{r_{\rm{a}}}}}} \right)}^{N - 1}},{r_{\min }} \le {r_{\rm{u}}} \le 2{r_{\rm{e}}} + {r_{\min }},}\\
{0,\qquad \qquad \qquad \qquad \qquad \quad {{\rm{otherwise}}}.}
\end{array}} \right.
\end{aligned}$}
\end{equation}
Then, we compared the approximated theoretical CDFs and PDFs of ${R_{\rm{d}}}$ with the empirical results through Monte Carlo simulations, as shown in Fig. \ref{CDF} and Fig. \ref{PDF}. It is shown that the proposed approximated values match well with the empirical results, verifying the effectiveness of the proposed approximation approach. %. when ${{r_{\min }}}$ is 600km, 1200km, and 1800km, respectively


\begin{comment}
\begin{figure*}[t] 
\textcolor{blue}{
\begin{equation}\label{cdfrd}
\resizebox{0.8\hsize}{!}{$
\begin{aligned}
{F_{{R_{\rm{d}}}}}\left( {{r_{\rm{d}}}} \right) = \left\{ {\begin{array}{*{20}{l}}
{1 - {{\left( {1 - \frac{{r_{\rm{d}}^{\rm{2}} - R_{{\rm{bd}}}^{\rm{2}} - r_{\min }^2}}{{4{r_{\rm{e}}}{r_{\rm{a}}}}}} \right)}^N},\sqrt {r_{\min }^2 + R_{{\rm{bd}}}^2}  \le {r_{\rm{d}}} \le \sqrt {{{\left( {2{r_{\rm{e}}} + {r_{\min }}} \right)}^2} + R_{{\rm{bd}}}^2} ,}\\
{0,\qquad \qquad \qquad \qquad \qquad {{\rm{otherwise}}}.}
\end{array}} \right.
\end{aligned}$}
\end{equation}}
\end{figure*}
%\textcolor{blue}{Accordingly, the PDF can be expressed in (\ref{pdfrd}) }
\begin{figure*}[t] 
\textcolor{blue}{
\begin{equation}\label{pdfrd}
\resizebox{0.8\hsize}{!}{$
\begin{aligned}
{f_{{R_{\rm{d}}}}}\left( {{r_{\rm{d}}}} \right) = \left\{ {\begin{array}{*{20}{l}}
{\frac{{{r_{\rm{d}}}N}}{{2{r_{\rm{e}}}{r_{\rm{a}}}}}{{\left( {1 - \frac{{r_{\rm{d}}^{\rm{2}} - R_{{\rm{bd}}}^2 - r_{\min }^2}}{{4{r_{\rm{e}}}{r_{\rm{a}}}}}} \right)}^{N - 1}},\sqrt {r_{\min }^2 + R_{{\rm{bd}}}^2}  \le {r_{\rm{d}}} \le \sqrt {{{\left( {2{r_{\rm{e}}} + {r_{\min }}} \right)}^2} + R_{{\rm{bd}}}^2} ,}\\
{0,\qquad \qquad \qquad \qquad \qquad \quad {{\rm{otherwise}}}.}
\end{array}} \right.
\end{aligned}$}
\end{equation}}
%\hrulefill
\end{figure*}
 \end{comment}
 
\subsection{End-to-end Transmission Success Probability}\label{AA}
In the marine link, the success probability of signals transmitted from the NSC to the DS through the BS is defined as the received SNR at the DS being above predefined threshold $\tau $ denoted by ${P_{{\rm{bd}}}}(\tau )$ and given by
\begin{equation}
{P_{{\rm{bd}}}}(\tau ) \buildrel \Delta \over = \mathbb{P}\left( {{\rm{SN}}{{\rm{R}}_{{\rm{bd}}}} > \tau } \right),
\end{equation}
\noindent where $\tau$ represents the minimum SNR required for successful signal transmission. In contrast, when ${\rm{SN}}{{\rm{R}}_{{\rm{bd}}}}$ is below the predefined threshold, the signal transmission should use the space link. Then, the success probability of signals transmitted from the ES to the SS in the first time slot, followed by the SS to the DS in the second time slot, is denoted by ${P_{{\rm{esd}}}}\left( \tau  \right)$, defined as both the SNR received at SS in the first time slot in (\ref{snru}) and the SINR received at DS in the second time slot in (\ref{sinrd}) being over predefined threshold $\tau $, which can be expressed as
\begin{equation}
{P_{{\rm{esd}}}}\left( \tau  \right) \buildrel \Delta \over =\mathbb{P} \left( {{\rm{SN}}{{\rm{R}}_{\rm{u}}}{\rm{ > }}\tau {\rm{, SIN}}{{\rm{R}}_{\rm{d}}} > \tau } \right).
\end{equation}

In the LEO-SSCN, we focus on the end-to-end transmission success probability for the NSC transmitting signals to the DS via both links. Following the communication scheme and the performance measure, the end-to-end transmission success probability is a function of the success probability in the marine link and the success probability in the space link under the condition that the SNR in the marine link is below the predefined threshold value.

\begin{thm}
\label{thm1}
The end-to-end transmission success probability of the LEO-SSCN for an arbitrarily located DS on the sea surface is
\begin{equation}
\begin{aligned}
{P_{\rm{s}}}(\tau ) &= {P_{{\rm{bd}}}}(\tau ) + \left( {1 - {P_{{\rm{bd}}}}\left( \tau  \right)} \right)  \times {P_{{\rm{esd}}}}(\tau ),\\
\end{aligned}
\end{equation}
\noindent where ${P_{{\rm{bd}}}}(\tau )$ can be expressed as
\begin{equation}\label{coveragelink1}
\begin{aligned}
{P_{{\rm{bd}}}}(\tau ) = {Q_1}\left( {\frac{\upsilon }{\rho },\frac{1}{\rho }\sqrt {\frac{{\tau \sigma _{{\rm{bd}}}^2R_{{\rm{bd}}}^{{\alpha _{{\rm{bd}}}}}}}{{{p_{{\rm{bd}}}}{G_{{\rm{bd}}}}}}} } \right),
\end{aligned}
\end{equation}
\noindent and ${P_{{\rm{esd}}}}(\tau )$ can be expressed in (\ref{coveragelink2}), shown at the top of the next page, where $s_1 = \frac{{\tau {{\left( {r_{\rm{u}}^2 + R_{{\rm{bd}}}^2} \right)}^{\frac{{{\alpha}}}{2}}}}}{{{p_{\rm{d}}}{G_{\rm{d}}}}}$, ${{{\cal L}_{\left( {{I_s} + \sigma _{\rm{d}}^{\rm{2}}} \right)}}\left( {s_1\left( {\beta  - \delta } \right)} \right)}$ is the Laplace transform of the sum power of the interfering satellites $I_s$ and the additive noise power $\sigma _{\rm{d}}^{\rm{2}}$ in the second time slot of the space link.

\begin{figure*}[t] 
\begin{equation}\label{coveragelink2}
\resizebox{0.9\hsize}{!}{$
\begin{aligned}
{P_{{\rm{esd}}}}\left( \tau  \right) {\rm{ = }} &  \int_{{r_{\min }}}^{{r_{\max }}} {\left( {\mu \sum\limits_{n = 0}^{m - 1} {\frac{{{{\left( {1 - m} \right)}_n}{{\left( { - \delta } \right)}^n}}}{{{{\left( {n!} \right)}^2}}}} \sum\limits_{l = 0}^n {\frac{{n!}}{{l!}}{{\left( {\beta  - \delta } \right)}^{ - \left( {n + 1 - l} \right)}}} {{\left( {\frac{{\tau \sigma _{\rm{u}}^{\rm{2}}r_{\rm{u}}^{{\alpha }}}}{{{p_{\rm{u}}}{G_{\rm{u}}}}}} \right)}^l}{{\rm{e}}^{ - \left( {\beta  - \delta } \right)\frac{{\tau \sigma _{\rm{u}}^{\rm{2}}r_{\rm{u}}^{{\alpha }}}}{{{p_{\rm{u}}}{G_{\rm{u}}}}}}}} \right)} \\ 
 & \times \left( {\mu \sum\limits_{n = 0}^{m - 1} {\frac{{{{\left( {1 - m} \right)}_n}{{\left( { - \delta } \right)}^n}}}{{{{\left( {n!} \right)}^2}}}} \sum\limits_{l = 0}^n {\frac{{n!}}{{l!}}} {{\left( {\beta  - \delta } \right)}^{ - \left( {n + 1} \right)}}{{\left( { - s_1} \right)}^l}\frac{{{\partial ^{(l)}}}}{{\partial {s_1^{\left( l \right)}}}}{{\cal L}_{\left( {I_s + \sigma _{{\rm{d}}}^{{\rm{2}}}} \right)}}\left( {s_1\left( {\beta  - \delta } \right)} \right)} \right) {\left( {1 - \frac{{r_{\rm{u}}^2 - r_{\min }^2}}{{4{r_{\rm{e}}}{r_a}}}} \right)^{N - 1}}\frac{{{r_{\rm{u}}}N}}{{2{r_{\rm{e}}}{r_a}}}d{r_{\rm{u}}}.\\
\end{aligned}$}
\end{equation}
\hrulefill%black line
\end{figure*}
%\noindent In (\ref{coveragelink2}),
\end{thm}

The \textit{Proof} of Theorem \ref{thm1} can be seen in Appendix A.

To simplify the calculation, the Marcum Q-function can be approximated as \cite{end1}
\begin{equation}
{Q_1}\left( {a,b} \right) \approx \exp \left( { - {e^{{v_1}\left( a \right)}}{b^{{\mu _1}\left( a \right)}}} \right),
\end{equation}
\noindent where ${\mu _1}\left( a \right) = 2.174 - 0.592a + 0.593{a^2} - 0.092{a^3} + 0.005{a^4}$ and ${v_1}\left( a \right) =  - 0.840 + 0.327a - 0.740{a^2} + 0.083{a^3} - 0.004{a^4}$. Then, (\ref{coveragelink1}) can be simplified to
\begin{equation}
\resizebox{0.95\hsize}{!}{$
\begin{aligned}
{Q_1}\left( {\frac{\upsilon }{\rho },\frac{1}{\rho }\sqrt {\frac{{\tau \sigma _{{\rm{bd}}}^2R_{{\rm{bd}}}^{{\alpha _{{\rm{bd}}}}}}}{{{p_{{\rm{bd}}}}{G_{{\rm{bd}}}}}}} } \right) \approx \exp \left( { - {e^{{v_1}\left( {\frac{v}{\rho }} \right)}}{{\left( {\frac{1}{\rho }\sqrt {\frac{{\tau \sigma _{{\rm{bd}}}^2R_{{\rm{bd}}}^{{\alpha _{{\rm{bd}}}}}}}{{{p_{{\rm{bd}}}}{G_{{\rm{bd}}}}}}} } \right)}^{{\mu _1}\left( {\frac{v}{\rho }} \right)}}} \right).
\end{aligned}$}
\end{equation}

\subsection{Average Transmission Rate Capacity}
In the marine link, the performance measure of the capacity is defined as the ergodic capacity from the Shannon-Hartley theorem, given by
\begin{equation}
C_{\rm{bd}} \buildrel \Delta \over = B_{\rm{bd}}\left[ {{{\log }_2}\left( {1 + {\rm{SNR_{\rm{bd}}}}} \right)} \right],
\end{equation}
\noindent where $B_{\rm{bd}}$ is the bandwidth of the marine link. Accordingly, the performance measure of the capacity in the space link is defined as
\begin{equation}
{C_{{\rm{esd}}}} \buildrel \Delta \over = {B_{{\rm{esd}}}}\left[ {{{\log }_2}\left( {1 + {\rm{SIN}}{{\rm{R}}_{{\rm{min}}}}} \right)} \right],
\end{equation}
\noindent where $B_{\rm{esd}}$ is the bandwidth of the space link,  ${\rm{SIN}}{{\rm{R}}_{{\rm{min}}}}$ presents the smaller value between the received SNR at the SS and the received SINR at the DS, i.e., ${\rm{SIN}}{{\rm{R}}_{\min }} = \min \left\{ {{\rm{SN}}{{\rm{R}}_{\rm{u}}},{\rm{ SIN}}{{\rm{R}}_{\rm{d}}}} \right\}$. 

In the LEO-SSCN, the average transmission rate capacity is regarded as the ergodic capacity for both the marine link and the space link from the Shannon-Hartley theorem under the aforementioned communication scheme and performance measure. To be specific, the average transmission rate capacity is defined as the sum of the capacity of the marine link under the condition that the successful transmission occurs in the marine link and that of the space link under the condition that the unsuccessful transmission occurs in the marine link.

\begin{thm}
\label{thm2}
The average transmission rate capacity of the LEO-SSCN for an arbitrarily located DS on the sea surface is 
\begin{equation}
\resizebox{0.63\hsize}{!}{$
\begin{aligned}
&{{C_{\rm{s}}}}  = \int_{\sqrt {\frac{{\tau \sigma _{{\rm{bd}}}^2R_{{\rm{bd}}}^{{\alpha _{{\rm{bd}}}}}}}{{{p_{{\rm{bd}}}}{G_{{\rm{bd}}}}}}} }^\infty  {{f_{{H_{{\rm{bd}}}}}}\left( x \right) {C_{{\rm{bd}}}}dx}\\
&\qquad+ \int_0^{\sqrt {\frac{{\tau \sigma _{{\rm{bd}}}^2R_{{\rm{bd}}}^{{\alpha _{{\rm{bd}}}}}}}{{{p_{{\rm{bd}}}}{G_{{\rm{bd}}}}}}} } {{f_{{H_{{\rm{bd}}}}}}\left( x \right){C_{{\rm{esd}}}}dx}, 
\end{aligned}$}
\end{equation}
\noindent where the first item can be unfolded as 
\begin{equation}\label{capacitydir}
\resizebox{0.73\hsize}{!}{$
\begin{aligned}
&\int_{\sqrt {\frac{{\tau \sigma _{{\rm{bd}}}^2R_{{\rm{bd}}}^{{\alpha _{{\rm{bd}}}}}}}{{{p_{{\rm{bd}}}}{G_{{\rm{bd}}}}}}} }^\infty  {{f_{{H_{{\rm{bd}}}}}}\left( x \right){C_{{\rm{bd}}}}dx}\\
&{ = {B_{{\rm{bd}}}}\int_{\sqrt {\frac{{\tau \sigma _{{\rm{bd}}}^2R_{{\rm{bd}}}^{{\alpha _{{\rm{bd}}}}}}}{{{p_{{\rm{bd}}}}{{\rm{G}}_{{\rm{bd}}}}}}} }^\infty  {{{\log }_2}\left( {1 + \frac{{{p_{{\rm{bd}}}}{G_{{\rm{bd}}}}R_{{\rm{bd}}}^{ - {\alpha _{{\rm{bd}}}}}{x^2}}}{{\sigma _{{\rm{bd}}}^2}}} \right)} }\\
&{{\quad} \times \frac{x}{{{\rho ^2}}}\exp \left( {\frac{{ - \left( {{x^2} + {v^2}} \right)}}{{2{\rho ^2}}}} \right){I_0}\left( {\frac{{xv}}{{{\rho ^2}}}} \right)dx},
\end{aligned}$}
\end{equation}

\noindent and the second item can be unfolded in (\ref{capacityre}), shown at the top of this page, where ${s_2} = \frac{{\left( {{2^t} - 1} \right){\left( {r_{\rm{u}}^{\rm{2}} + R_{{\rm{bd}}}^2} \right)^{\frac{\alpha }{2}}}}}{{{p_{\rm{d}}}{G_{\rm{d}}}}}$ and ${{{\cal L}_{\left( {\sigma _{\rm{d}}^2 + I_{\rm{s}}} \right)}}\left( {s_2\left( {\beta  - \delta } \right)} \right)}$ presents the Laplace transform of the sum power of the interfering satellites $I_{\rm{s}}$ and received noise power $\sigma _{\rm{d}}^2$ in the second time slot of the space link, which can be obtained by replacing ${s_1}$ with ${s_2}$ in (${\ref{coveragelink2}}$). 

\begin{figure*}[t]
\begin{equation}\label{capacityre}
\resizebox{0.9\hsize}{!}{$
\begin{aligned}
\int_0^{\sqrt {\frac{{\tau \sigma _{{\rm{bd}}}^2R_{{\rm{bd}}}^{{\alpha _{{\rm{bd}}}}}}}{{{p_{{\rm{bd}}}}{G_{{\rm{bd}}}}}}} } {{f_{{H_{{\rm{bd}}}}}}\left( x \right){C_{{\rm{esd}}}}dx} =& {B_{{\rm{esd}}}}\int_0^{\sqrt {\frac{{\tau \sigma _{{\rm{bd}}}^2R_{{\rm{bd}}}^{{\alpha _{{\rm{bd}}}}}}}{{{p_{{\rm{bd}}}}{G_{{\rm{bd}}}}}}} }\int_{{r_{\min }}}^{{r_{\max }}} \int_{0}^{\infty} {\mu \mathop \sum \limits_{n = 0}^{m - 1} \frac{{{{\left( {1 - m} \right)}_n}{{( - \delta )}^n}}}{{{{(n!)}^2}}}\sum\limits_{l = 0}^n {\frac{{n!}}{{l!}}{{\left( {\beta  - \delta } \right)}^{ - (n + 1)}}{{\left( { - {s_2}} \right)}^l}} }\frac{{{\partial ^{(l)}}}}{{\partial s_2^{\left( l \right)}}}  \\
&\times {{\cal L}_{\left( {\sigma _{\rm{d}}^2 + I_{\rm{s}}} \right)}}\left( {{s_2}\left( {\beta  - \delta } \right)} \right)dt{\left( {1 - \frac{{r_{\rm{u}}^2 - r_{\min }^2}}{{4{r_{\rm{e}}}{r_a}}}} \right)^{N - 1}}\frac{{{r_{\rm{u}}}N}}{{2{r_{\rm{e}}}{r_a}}}d{r_{\rm{u}}}{\frac{x}{{{\rho ^2}}}\exp \left( {\frac{{ - \left( {{x^2} + {v^2}} \right)}}{{2{\rho ^2}}}} \right){I_0}\left( {\frac{{xv}}{{{\rho ^2}}}} \right)dx},
\end{aligned}$}
\end{equation}
\hrulefill%black line
\end{figure*}
\end{thm}

The \textit{Proof} of Theorem \ref{thm2} can be seen in Appendix B.

It is important to emphasize that both (\ref{coveragelink2}) and (\ref{capacityre}) cannot be straightforwardly expressed in closed form. The difficulty arises from combining the SR fading model with the complex CDF expression and the cumulative interference power of ISs calculations involving multiple variables. Furthermore, the end-to-end signal transmission encompassing both uplink and downlink channels in the space link further adds to the challenge of
simplifying these expressions.

Moreover, to gain more insights into Theorem 2, we investigate a special setting to simplify the general-case expression. When the Rician K factor of the Rician fading model in the marine link approaches infinity, denoted as ${K_{{\rm{Rician}}}} \to \infty $, the channel of the marine link only has the LoS component \cite{digital1, k-infi}. Therefore, the channel in the marine link degrades to an additive white Gaussian noise (AWGN) channel. This allows the capacity in the marine link to be approximated as
\begin{equation}
\label{K=00}
\begin{aligned}
   {C_{{\rm{bd, }}{K_{{\rm{Rician}}}} \to \infty }}   \approx   {B_{{\rm{bd}}}}{\log _2}\left( {1 + \frac{{{p_{{\rm{bd}}}}{G_{{\rm{bd}}}}R_{{\rm{bd}}}^{ - {\alpha _{{\rm{bd}}}}}}}{{\sigma _{{\rm{bd}}}^2}}} \right).
\end{aligned}
\end{equation}

When the distance from the BS to the DS is sufficiently short, satisfying the condition that the received SNR through the marine link at the DS is higher than the predefined threshold, signals are transmitted via the marine link. As a result, the average transmission rate capacity can be represented by the capacity of the marine link. Otherwise, signals should be transmitted through the space link, and the average transmission rate capacity is replaced by the capacity of the space link. Therefore, the average transmission rate capacity can be expressed as 
\begin{equation}
\label{K=002}
\begin{aligned}
   {C_{\rm{s}}} = \left\{ {\begin{array}{*{20}{l}}
{{C_{{\rm{bd, }}{K_{{\rm{Rician}}}} \to \infty }}, \quad {R_{{\rm{bd}}}} \le {{\left( {\frac{{{p_{{\rm{bd}}}}{G_{{\rm{bd}}}}}}{{\tau \sigma _{{\rm{bd}}}^2}}} \right)}^{\frac{1}{{{\alpha _{{\rm{bd}}}}}}}},}\\
{{C_{{\rm{esd}}}}, \qquad \qquad \qquad  {\rm{otherwise}}{\rm{.}}}
\end{array}} \right.
\end{aligned}
\end{equation}

\begin{comment}
\textcolor{blue}{\emph{Case 1:} When the distance from the BS to the DS is sufficiently short, satisfying the condition that the received SNR through the marine link at the DS is higher than the predefined threshold, the majority of signals are transmitted via the marine link. As a result, the average transmission rate capacity can be approximated by the capacity of the marine link, which can be expressed as }
\begin{equation}
\label{r=0}
\begin{aligned}
  \textcolor{blue}{ {C_{\rm{s}}} \approx {C_{{\rm{bd, }}{K_{{\rm{Rician}}}} \to \infty }},\quad{R_{{\rm{bd}}}} < {\left( {\frac{{{p_{{\rm{bd}}}}{G_{{\rm{bd}}}}}}{{\tau \sigma _{{\rm{bd}}}^2}}} \right)^{\frac{1}{{{\alpha _{{\rm{bd}}}}}}}}.}
\end{aligned}
\end{equation}

\textcolor{blue}{\emph{Case 2:} When the distance from the BS to the DS is long enough, satisfying the condition that the received SNR at the DS is lower than the predefined threshold, most of the signals are transmitted through the space link. In this case, the average transmission rate capacity can be approximately replaced by the capacity of the space link, which can be expressed as} 
\begin{equation}
\label{r=00}
\begin{aligned}
   \textcolor{blue}{{C_{\rm{s}}} \approx {C_{{\rm{esd}}}},\quad{R_{{\rm{bd}}}} \ge {\left( {\frac{{{p_{{\rm{bd}}}}{G_{{\rm{bd}}}}}}{{\tau \sigma _{{\rm{bd}}}^2}}} \right)^{\frac{1}{{{\alpha _{{\rm{bd}}}}}}}}.}
\end{aligned}
\end{equation}
\end{comment}

\section{Numerical Results}
In this section, we validate our derivations for the end-to-end transmission success probability and average transmission rate capacity through Monte Carlo simulations. Moreover, we provide numerical results to study the impact of different network parameters on the end-to-end signal transmission performance, including constellation altitude, constellation size, the number of orthogonal frequency channels in the space link, and the distance from the BS to the DS. Then, we examine the impact of various parameters, such as the predefined SNR threshold and the distance from the BS to the DS, on the performance improvement of the LEO-assisted end-to-end transmission compared with the marine link, including the increment of transmission success probability and the increment of transmission rate capacity. They provide important guidelines to the LEO-SSCN for signal transmissions from shore to ship.

\subsection{Parameter Settings}
In the simulation, NSC is placed on the Earth at point $\left({0,0,{r_{\rm{e}}}} \right)$. The DS is randomly generated on the sea surface with a distance ${R_{{\rm{bd}}}}$ from the BS. The $N$ satellite points are generated on a spherical surface at constellation altitude ${r_{\min }}$ above the Earth's surface, forming a BPP. Unless otherwise specified, $N$ is set to 1000, ${r_{\min }}$ is set to 1200 km, ${R_{{\rm{bd}}}}$ is set to 40 nautical miles (n miles), $K$ is set to 10 and predefined threshold $\tau$ is set to 10 dB.
\begin{table}[t]
	\caption{Parameter Settings}
	\label{Parameter value Settings}
	\centering
	\begin{tabular}{c|c}
		\hline
		\hline
		Parameters & Values \\ \hline
		$ {r_{\rm{e}}};  {K_{{\rm{Rician}}}} $	&  6371 km; 10    \\ \hline
		${p_{\rm{bd}}};{p_{\rm{u}}};{p_{\rm{d}}};{p_{{\rm{i}}}}$  & 80 W; 25 W; 10 W; 10 W\\ \hline
        ${G_{{\rm{bd}}}};{G_{\rm{u}}};{G_{\rm{d}}};{G_{{\rm{i}}}}$  & 2 dBi;  48 dBi; 38.5 dBi; 28.5 dBi\\ \hline
		${\alpha _{{\rm{bd}}}};{\alpha}$ & $2.9;2.4$ \\ \hline
		$\sigma _{{\rm{bd}}}^2;\sigma _{\rm{u}}^2;\sigma _{\rm{d}}^2$		&  $-$100 dBm; $-$90 dBm; $-$90 dBm    \\ \hline
		$ SR\left( {{b},m,\Omega } \right)$	 &   $SR\left( {0.3,{\rm{ }}3,{\rm{ }}0.4} \right)$     \\ \hline
		$\kappa{\rm{ - }}\mu \left( {\kappa,\mu ,m} \right)$ & $k{\rm{ - }}\mu \left( {2/3,{\rm{ }}1,{\rm{ }}3} \right)$\\ \hline
			${B_{{\rm{bd}}}};{\rm{ }}{B_{{\rm{esd}}}}$ & 30 MHz; 250 MHz\\ \hline
		\hline
	\end{tabular}
\end{table}

For the marine link, the transmit power and the antenna gain of the BS are set as 80 W and 2 dBi \cite{pd1}, respectively. For the space link, the transmit power and the antenna gain of the ES are set to be 25 W and 48 dBi \cite{48dB}, respectively. The SS is set with a transmit power of 10 W and an antenna gain of 38.5 dBi, while the IS is set with a transmit power of 10 W and an antenna gain of 28.5 dBi, respectively \cite{d3, d10}. When the parameters in the SR fading model $\left( {{b},m,\Omega } \right)$ are set as $\left(0.3, 3, 0.4\right)$, correspondingly, the parameters in the $\kappa  - \mu$ fading model $ \left( {\kappa ,\mu ,m} \right)$ are calculated as $\left(2/3, 1, 3\right)$. The noise power spectral densities (PSDs) ${N_0}$ are set to $-$174 dBm/Hz in both the marine link \cite{pd4} and the space link \cite{d9, d10}. The bandwidth of the channel in the marine link and the bandwidth of each channel in the space link are set as 30 MHz and 250 MHz \cite{pd1, pr3}, respectively. Therefore, by applying the relationship among noise PSD, bandwidth $B$, and noise power ${\sigma ^2}$, that is ${N_0}B = {\sigma ^2}$ \cite{n0b}, the noise power is $-$100 dBm in the marine link and $-$90 dBm in the space link. Other parameters used in the simulations are summarized in Table \ref{Parameter value Settings}.  

\begin{figure}[t]
	\centerline{\includegraphics[scale=0.55]{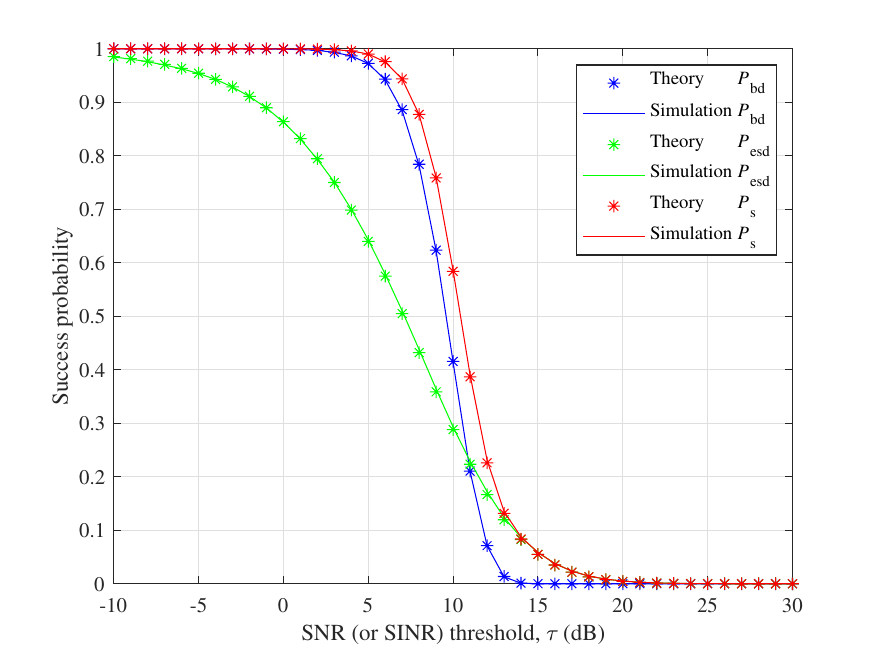}}%cov-T20240204
	\caption{Comparison of theory and simulation of the success probability in different links.}
	\label{cov-link}
\end{figure}

% and $-$203 dBm/Hz \cite{-203}, respectively The noise power spectral density (PSD) ${N_0}$ for the marine link and the Ka-band communication in the space link are set to $-$174 dBm/Hz and $-$203dBm/Hz \footnote{Within our model, there is no strict requirement for the PSDs of the two links to be identical. The PSD in the spack link can also be set to other values, such as -203dBm/Hz \cite{-203}.}

%For the marine link, the transmit power and the antenna gain of the BS are set as 80 w and 2 dBi \cite{pd1}, respectively. The noise power spectral density (PSD) for the marine link is -174 dBm/Hz \cite{pd4}. For the space link, the transmit power and the antenna gain of the ES are set to be 50 w and 2 dBi, respectively. The SS is set with a transmit power of 10 w and an antenna gain of 6 dBi \cite{d3, pr1}. The noise PSD for the Ku-band communication in the space link is -203 dBm/Hz \cite{-203}. When the parameters in the SR fading model $\left( {{b},m,\Omega } \right)$ are set as $\left(0.3, 3, 0.4\right)$, correspondingly, the parameters in the $\kappa  - \mu$ fading model $ \left( {\kappa ,\mu ,m} \right)$ is calculated as $\left(2/3, 1, 3\right)$. The bandwidth of the channel in the marine link and the bandwidth of each channel in the space link are set as 30 MHz and 250 MHz, respectively \cite{pd1, pr3}.  Other parameters used in the simulations are summarized in Table \ref{Parameter value Settings}. 
\subsection{Results and Discussion}
\begin{figure}[t]
		\centerline{\includegraphics[scale=0.55]{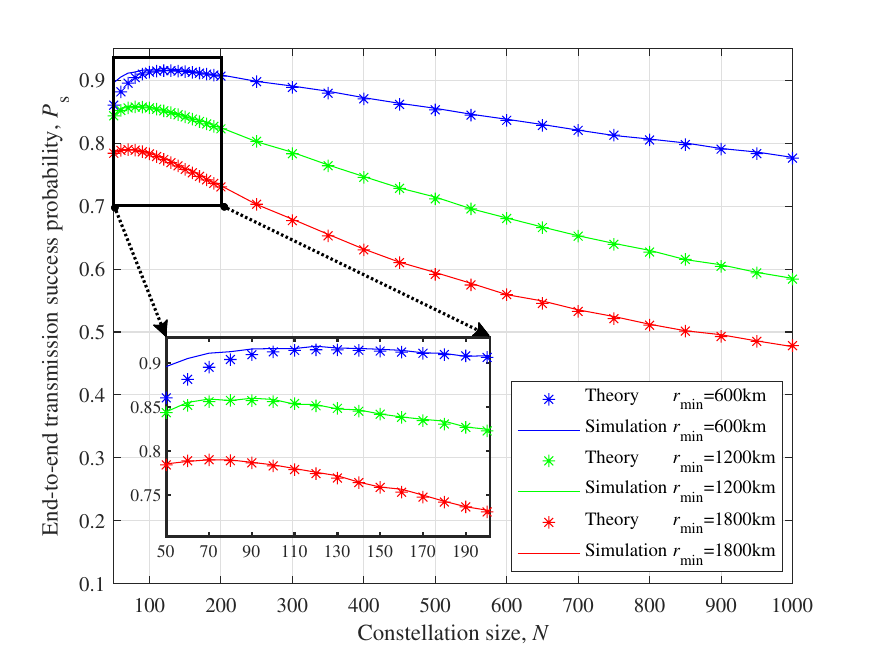}}%cov-N20240204
		\caption{Effect of the constellation size on the end-to-end transmission success probability.}
		\label{cov-N}
 \end{figure}

\begin{figure}[t]
		\centerline{\includegraphics[scale=0.55]{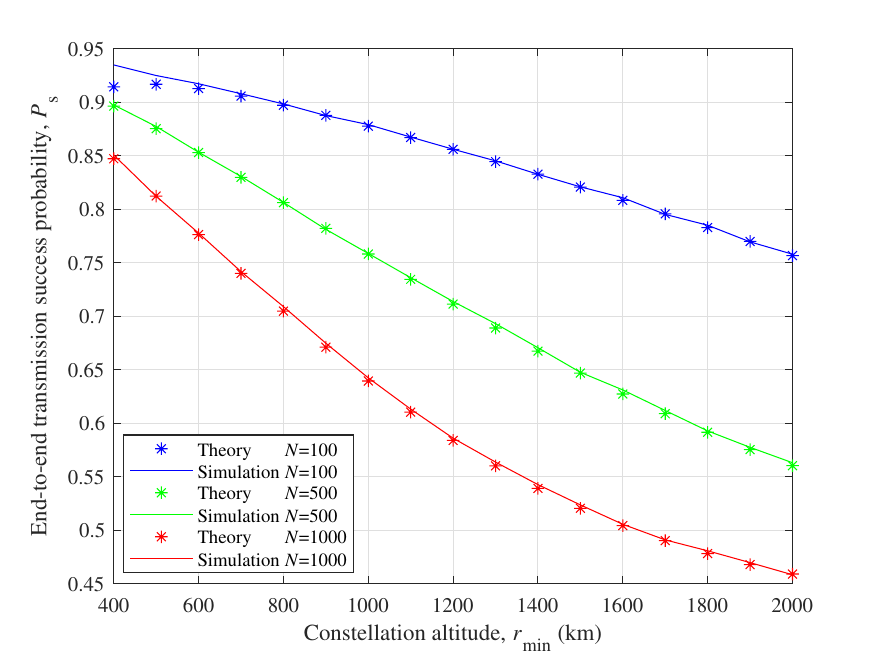}}%cov-rmin20240204
		\caption{Effect of the constellation altitude on the end-to-end transmission success probability.}
		\label{cov-rmin}
 \end{figure}

\begin{figure}[t]
		\centerline{\includegraphics[scale=0.55]{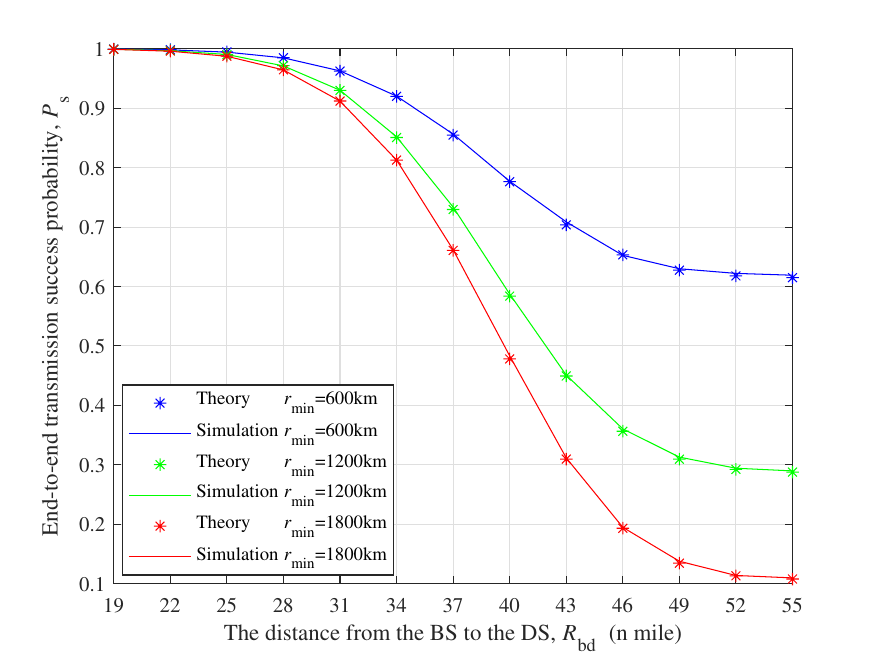}}	%cov-rbv20240204
		\caption{Effect of the distance from the BS to the DS on the end-to-end transmission success probability.}
		\label{cov-rbv}
 \end{figure}

\begin{figure}[t]
		\centerline{\includegraphics[scale=0.55]{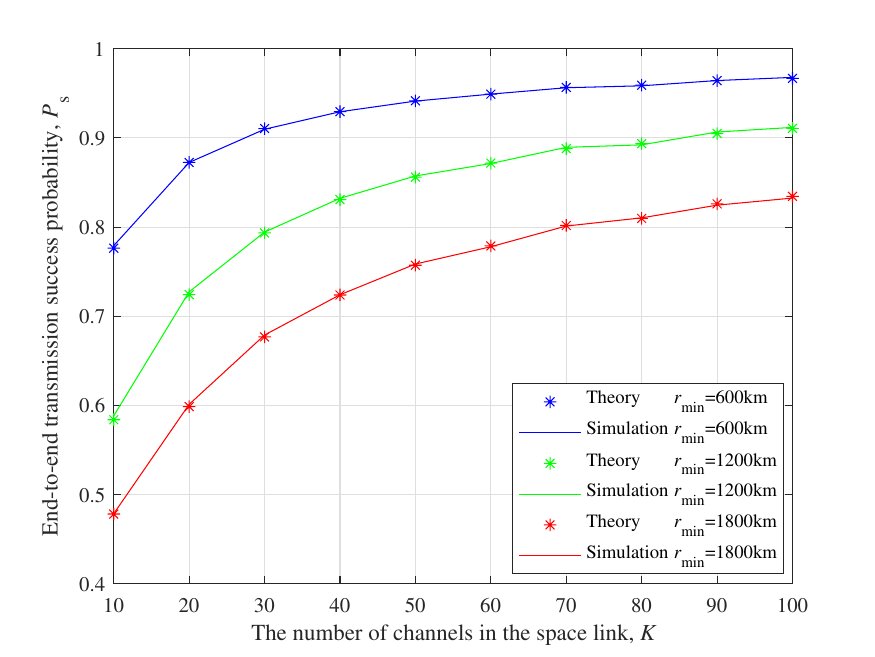}}%cov-K20240204	
		\caption{Effect of the number of channels in the space link on the end-to-end transmission success probability.}
		\label{cov-K}
 \end{figure}

In all the figures, markers represent the derived analytical results while the solid lines represent the Monte Carlo simulations. We validate our expressions for the end-to-end transmission success probability derived in Theorem \ref{thm1}, as shown from Fig. \ref{cov-link} to Fig. \ref{cov-K}. Fig. \ref{cov-link} demonstrates the success probability varies with a predefined SNR (or SINR) threshold $\tau$ from -10 dB to 30 dB in different links, including the success probability in the marine link ${P_{{\rm{bd}}}}$, the success probability in the space link ${P_{{\rm{esd}}}}$ and the end-to-end transmission success probability ${P_{{\rm{s}}}}$, respectively. The theoretical results match the simulation results indicating the accuracy of the derivations. The success probability in the marine link depends on the distance from the BS to the DS, whereas that in the space link depends on the constellation altitude, constellation size, and the number of orthogonal frequency channels. In addition, when the SNR (or SINR) threshold is less than 11 dB, the success probability in the marine link is greater than that in the space link. In this case, the NSC tends to choose the marine link for signal transmissions. It can be observed that when the SNR (or SINR) threshold is 8 dB, the success probability in the marine link is about 0.788, while by incorporating the space link, the end-to-end transmission success probability ${P_{{\rm{s}}}}$ is about 0.881, achieving the growth rate of success probability 12${\rm{\% }}$. However, when the SNR (or SINR) threshold is larger than 11 dB, the space link outperforms the marine link. Consequently, the NSC prefers the space link. For instance, when the SNR (or SINR) threshold is 13 dB, the success probability in the marine link is about 0.014, while by incorporating the space link, the end-to-end transmission success probability ${P_{{\rm{s}}}}$ is 0.138, achieving an increment of success probability 886${\rm{\% }}$. This superior performance is attributed to the fact that the space link uses a wider bandwidth and greater power for signal transmission compared to the maritime link.  It’s undeniable that the integration of the space link inevitably incurs additional expenses.

\begin{figure}[t]
		\centerline{\includegraphics[scale=0.55]{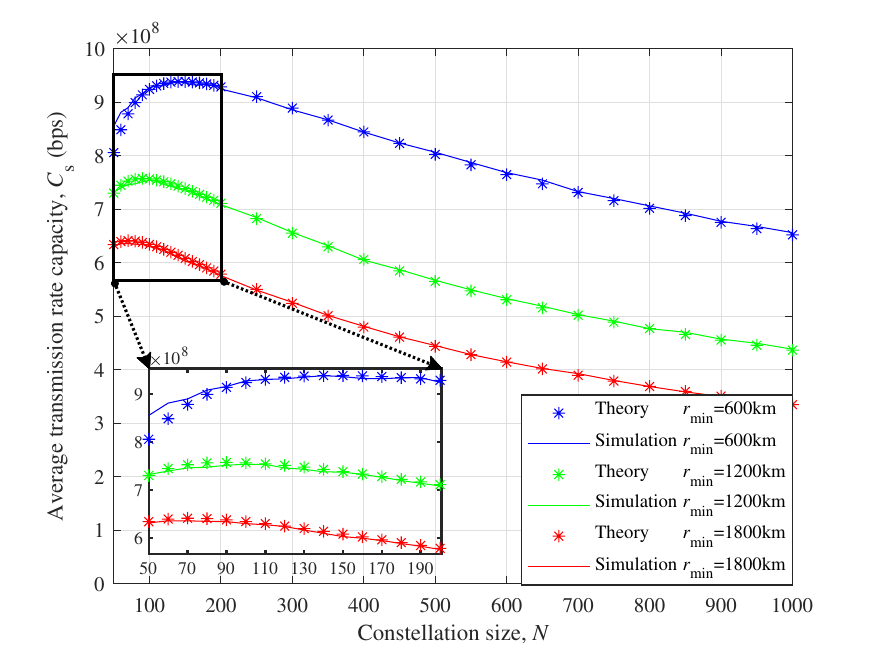}}%	cap-N20240204-2
		\caption{Effect of the constellation size on average transmission rate capacity.}
		\label{cap-N}
 \end{figure}

Fig. \ref{cov-N} shows the relationship between the end-to-end transmission success probability and the constellation size when the constellation altitude is set to 600 km, 1200 km, and 1800 km, respectively. It can be seen that as the constellation size increases, there is an initial surge in the end-to-end transmission success probability, reaching a peak. For instance, when the constellation altitude is set at 600 km, the highest end-to-end transmission success probability reaches 0.915, reaching its pinnacle at a constellation size of 120. This is because the increase in constellation size results in a heightened density of LEO satellites at a given altitude. Consequently, the SS may be closer to the point $\left( {0,0,{r_{\rm{e}}}} \right)$, bringing a shortened distance of both ${R_{\rm{u}}}$ and ${R_{\rm{d}}}$. The decreased distances from the ES to the SS and from the SS to the DS contribute to a gradual improvement in space link quality. During this ascending phase, the improvement in link quality becomes more noticeable than the impact of interfering satellites. As the constellation size continues to grow, the end-to-end transmission success probability experiences a subsequent decline. This is because a larger number of satellites results in an increased number of interfering satellites being assigned to the same channel, leading to higher interference power in the downlink space link. Therefore, the optimal constellation size falls within the range of 80 to 120. This also explains the similar trend illustrated in Fig. \ref{cap-N} for the average transmission rate capacity. It is noteworthy that a disparity emerges between theory and simulation when the number of satellites is less than 100 at a constellation altitude of 600 km. This discrepancy primarily stems from the fact that a smaller constellation size at a lower constellation altitude results in the shape of the constructed triangle deviating significantly from that of a right triangle, leading to substantial errors in the proposed distance approximation approach for the distance from the SS to the DS.

 \begin{figure}[t]
		\centerline{\includegraphics[scale=0.55]{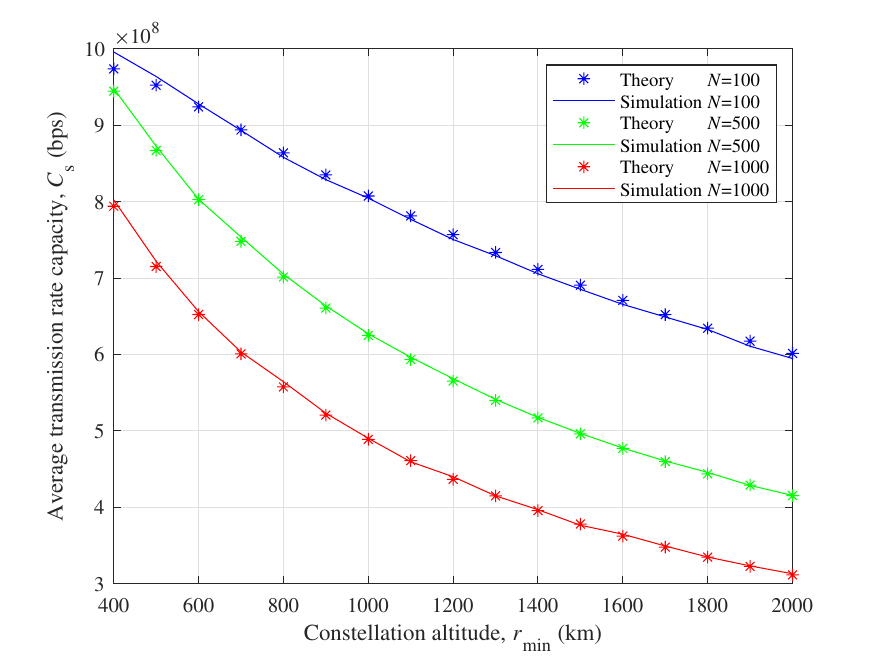}}%cap-rmin20240204
		\caption{Effect of the constellation altitude on average transmission rate capacity.}
		\label{cap-rmin}
 \end{figure}
 
  \begin{figure}[t]
		\centerline{\includegraphics[scale=0.55]{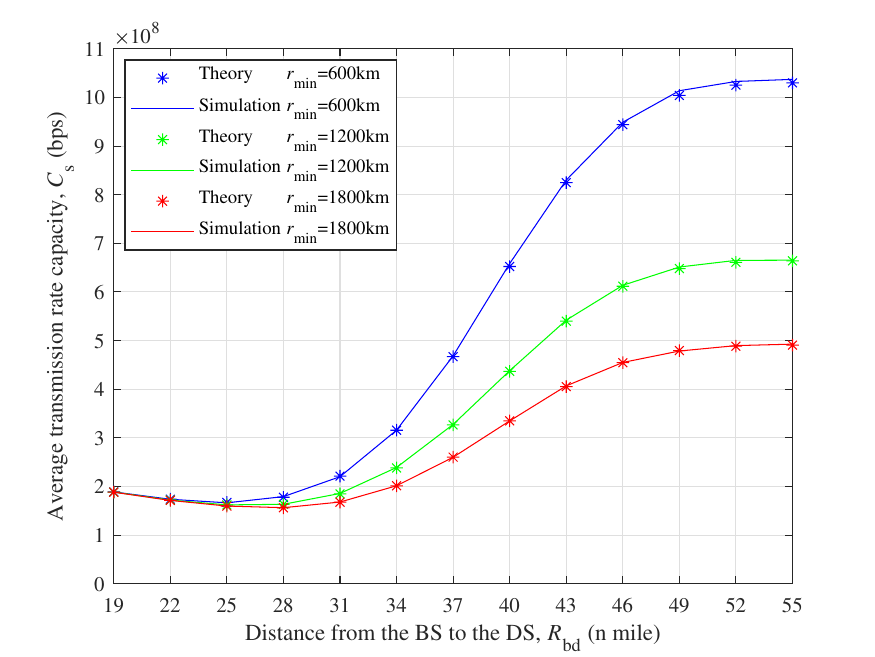}}	%cap-rbv20240204
		\caption{Effect of the distance from the BS to the DS on average transmission rate capacity.}
		\label{cap-rbv}
 \end{figure}
  \begin{figure}[t]
		\centerline{\includegraphics[scale=0.55]{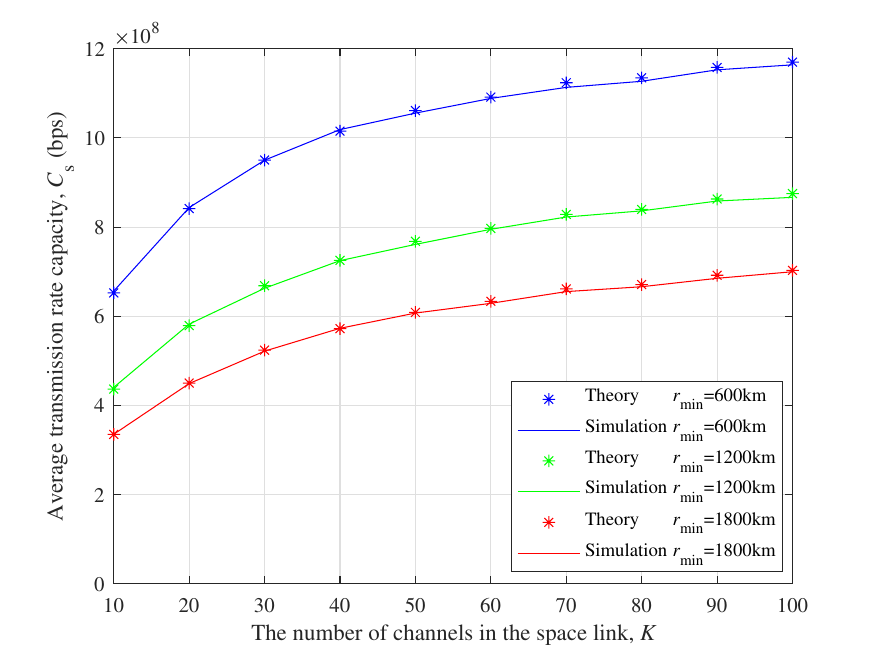}}%cap-K20240204
		\caption{Effect of the number of channels in the space link on average transmission rate capacity.}
		\label{cap-K}
 \end{figure}
\begin{comment}
    \begin{figure*}[t]
	\centering
	\begin{minipage}{0.49\linewidth}
		\centering
		\includegraphics[width=0.9\linewidth]{cap-rmin630.pdf}
		\caption{Effect of the constellation altitude on average transmission rate capacity.}
		\label{cap-rmin}
	\end{minipage}
	\begin{minipage}{0.49\linewidth}
		\centering
		\includegraphics[width=0.9\linewidth]{cap-N630.pdf}
		\caption{Effect of the constellation size on average transmission rate capacity.}
		\label{cap-N}
	\end{minipage}
%\qquad
 \begin{minipage}{0.49\linewidth}
		\centering
		\includegraphics[width=0.9\linewidth]{cap-rbv630.pdf}
		\caption{Effect of the distance from the BS to the DS on average transmission rate capacity.}
		\label{cap-rbv}
	\end{minipage}
\begin{minipage}{0.49\linewidth}
		\centering
		\includegraphics[width=0.9\linewidth]{cap-k630.pdf}
		\caption{Effect of the number of channels in the space link on average transmission rate capacity.}
		\label{cap-K}
	\end{minipage}
 	%\qquad
\end{figure*}
\end{comment}

Several guidelines for the LEO-SSCN can be deduced from Fig. \ref{cov-rmin}. It shows the relationship between the end-to-end transmission success probability and the constellation altitude when the constellation size is set at 100, 500, and 1000, respectively. The results indicate that as the constellation altitude increases, the end-to-end transmission success probability decreases. This is due to the fact that the increasing constellation altitude results in a long distance from the ES to the SS, as well as from the SS to the DS, which increases the impact of path loss on signal transmissions. As a consequence, both the SNR received at the SS and the SINR received at the DS decrease, leading to a decline in the end-to-end transmission success probability. 
%In brief, Fig. \ref{cov-N} and Fig. \ref{cov-rmin} suggest that, for optimal performance, the constellation altitude and constellation size should be carefully considered in the design of the LEO-SSCN.

Fig. \ref{cov-rbv} examines the impact of the distance from the BS to the DS on the end-to-end transmission success probability. It can be observed that, as ${R_{{\rm{bd}}}}$ increases, the end-to-end transmission success probability remains stable first, then it drops and finally becomes constant. The NSC tends to choose the marine link for signal transmissions when the ${R_{{\rm{bd}}}}$ is small enough that the SNR received by the DS through the marine link can meet the threshold requirement. As ${R_{{\rm{bd}}}}$ gradually increases, the path loss in the marine link becomes more significant, resulting in a gradual decrease in the end-to-end transmission success probability. Eventually, when the SNR received at DS in the marine link decreases to the point where most signal transmissions cannot meet the SNR (or SINR) threshold requirements, the NSC has to switch to the space link for signal transmissions. In this case, compared with the distance between the ES and SS as well as the distance between the SS and the DS, ${R_{{\rm{bd}}}}$ is relatively small. This results in little influence on signal transmissions despite a further increased ${R_{{\rm{bd}}}}$ value.

 \begin{figure}[t]
		\centerline{\includegraphics[scale=0.55]{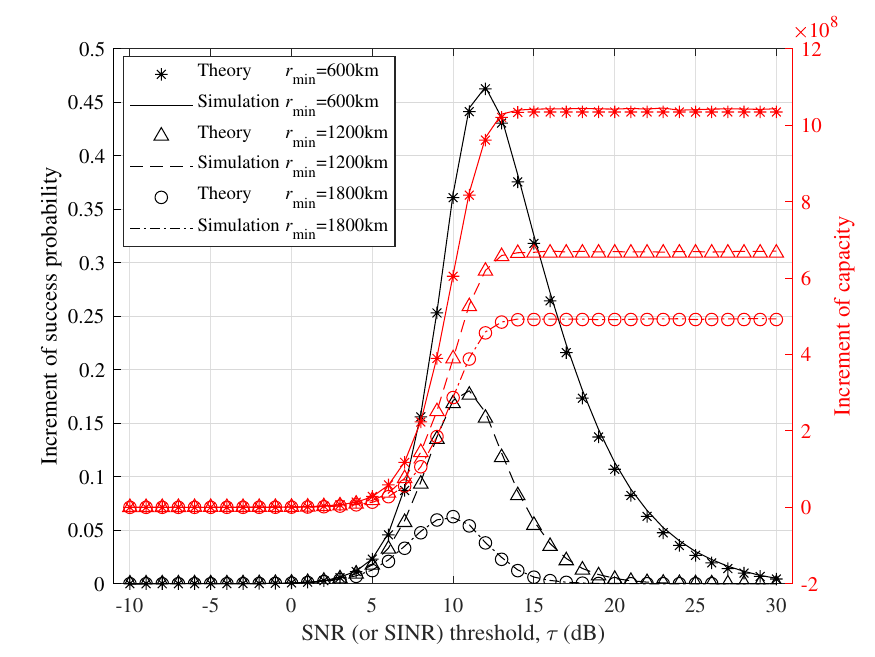}}	%incre-T20240204
		\caption{Increments of transmission success probability and transmission rate capacity with different predefined SNR (or SINR) threshold $\tau$.}
		\label{cov-incre-T}
 \end{figure}

Fig. \ref{cov-K} shows the relationship between the end-to-end transmission success probability and the number of channels in the space link. The results show that as the number of channels increases, the end-to-end transmission success probability first increases and then stays stable. This can be attributed to the reduction in the number of interfering satellites as the number of channels increases. To be specific, when the number of channels is less, a large number of interfering satellites are allocated in the same channel, resulting in a higher interference power. However, as the number of channels increases, the number of interfering satellites in the same channel decreases, leading to a reduction of the interference power. This, in turn, leads to an increase in the received SINR at the DS. When the number of channels in the space link is large, each channel is usually assigned only a few ISs or no ISs, leading to a constant SNR (or SINR) received at the DS and a stable end-to-end transmission success probability. In short, the number of channels in the space link has a significant impact on the end-to-end transmission success probability in the LEO-SSCN, and the increased number of channels can lead to improved performance.

Our derivations of the average transmission rate capacity, as stated in Theorem \ref{thm2}, are validated through Monte Carlo simulations presented in Figs. \ref{cap-rmin}$-$\ref{cap-K}. The average transmission rate capacity of the LEO-SSCN in relation to the constellation altitude and constellation size is presented in Fig. \ref{cap-rmin} and Fig. \ref{cap-N}, respectively. The results of Monte Carlo simulations align well with the theoretical expressions. Fig. \ref{cap-rmin} demonstrates average transmission rate capacity decreases with an increase in constellation altitude. This decrease can be attributed to two main factors. Firstly, a higher constellation altitude leads to a lower SINR at the DS due to the amplified influence of path loss. Secondly, the number of visible satellites increases as the constellation altitude increases, thus generating more ISs. Consequently, the combined effect of the increased path loss and a greater number of visible satellites adversely impacts the SINR and ultimately reduces the average transmission rate capacity of the LEO-SSCN. 

%\textcolor{blue}{Similarly, as illustrated in Fig. \ref{cap-N}, there is an initial increase in the average transmission rate capacity with the growing constellation size, reaching a peak value, and subsequently exhibiting a decline as the constellation size further increases. This trend can be attributed to the dominant role played by the enhancement of available link quality during the phase of rising average transmission rate capacity, in contrast to the influence of interfering satellites.} Conversely, during the phase of declining average transmission rate capacity, more satellites are allocated in each channel, leading to an increase in interference levels. Consequently, the increased ISs adversely affect the average transmission rate capacity of the LEO-SSCN. \textcolor{blue}{Therefore, the optimal constellation size occurs in the range from 80 to 120.}

%Similarly, as depicted in Fig. \ref{cap-N}, the average transmission rate capacity exhibits a decreasing trend as the constellation size increases. The reason is that as the constellation size grows, more satellites are allocated in each channel, leading to an increase in interference levels. Consequently, the increased ISs adversely affect the average transmission rate capacity of the LEO-SSCN.

The influence of the distance from the BS to the DS on the average transmission rate capacity is depicted in Fig. \ref{cap-rbv}. When ${R_{{\rm{bd}}}}$ is less than 28 n miles, the average transmission rate capacity gradually decreases as ${R_{{\rm{bd}}}}$ increases. This can be attributed to the fact that the marine link is preferred for signal transmissions in such cases, as both links can meet the signal transmission requirements. However, when ${R_{{\rm{bd}}}}$ exceeds 28 n miles, the average transmission rate capacity increases rapidly to its maximum value. This sudden surge can be attributed to the fact that the obtained SNR at the DS through the marine link falls below the required threshold value for a successful transmission, leading to a switch to the space link for better capacity performance. Thus, a surged average transmission rate capacity appears. 

Fig. \ref{cap-K} depicts the effect of the number of channels in the space link on average transmission rate capacity. As the number of orthogonal frequency channels in the space link increases from 10 to 100, we observe a gradual increase in the average transmission rate capacity. A little disparity is observed when ${r_{\min }}$ is 600 km compared to cases when ${r_{\min }}$ takes values of 1200 km and 1800 km in Figs. \ref{cap-rmin}$-$\ref{cap-K}. This can be attributed to the fact that a smaller constellation size at a lower constellation altitude results in the shape of the constructed triangle deviating significantly from that of a right triangle, leading to substantial errors in the proposed distance approximation approach for the distance from the SS to the DS. Furthermore, it is important to highlight that a disparity can arise when the antenna gain of the uplink is not significantly higher than that in the downlink. In such cases, the capacity of the uplink is often smaller than that of the downlink in many realizations of Monte Carlo simulations. This leads to the capacity of the space link being represented by the uplink, which deviates from the approximation approach that calculates capacity in the space link using the capacity of the downlink.

\begin{figure}[t]
		\centerline{\includegraphics[scale=0.55]{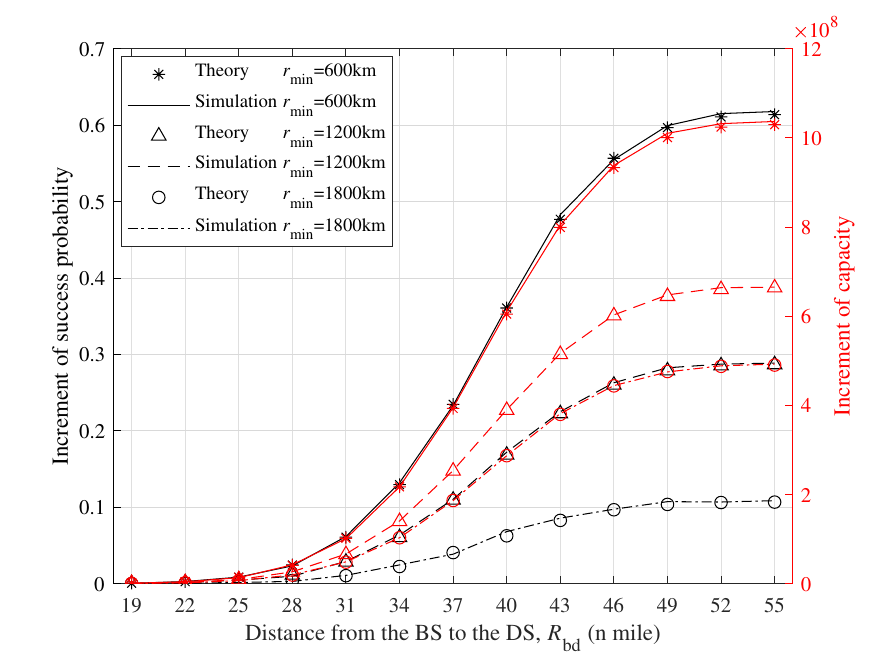}}	%incre-rbv20240204
		\caption{Effect of the distance from BS to DS on the increments of transmission success probability and transmission rate capacity.}
	    \label{biax-cre}
 \end{figure}

%However, it is noteworthy that when the number of channels exceeds 40, a discrepancy between the theory and simulation appears. The disparity primarily arises from the reduction in the number of ISs as the number of orthogonal frequency channels increases. To be specific, a large number of orthogonal frequency channels in the space link lead to a fluctuation in the number of ISs during each iteration of the Monte Carlo simulations.

The impact of the predefined SNR (or SINR) threshold $\tau$ on the increments of transmission success probability and transmission rate capacity under different constellation altitudes is presented in Fig. \ref{cov-incre-T}. The increment of transmission success probability is defined as the difference between the LEO-assisted end-to-end transmission success probability and the success probability in the marine link. It first exhibits a rise as the predefined threshold increases, peaking at 11 dB, and then decreases when the threshold further increases. The increment of transmission rate capacity is defined as the difference between the LEO-assisted average capacity and capacity in the marine link. When the threshold is below 5 dB, the value remains constant at 0. However, as the threshold further increases, it abruptly surges from 0 to reach a maximum value and then remains stable. These behaviors can be attributed to the fact that when the predefined threshold is small, the primary signal transmission occurs through the marine link. As the predefined threshold increases, the LEO-assisted end-to-end transmission gradually shifts to the space link, thereby enhancing these increments. However, as the predefined threshold continues to increase, both the marine link and the space link fail to meet the transmission requirements. Consequently, the increment of transmission success probability diminishes, and the increment of transmission rate capacity remains stable. As the constellation altitude decreases from 1800 km to 600 km, both the two increments significantly increase. This is because a lower constellation altitude results in shorter distances between the ES and the SS, as well as between the SS and the DS. Consequently, path loss in signal transmissions is significantly reduced, leading to an increased transmission success probability and rate capacity. From the results, it can be seen that relying solely on either the marine link or the space link fails to yield superior transmission performance. Therefore, employing LEO satellite-aided shore-to-ship communications becomes imperative, enabling dynamic adjusting to the transmission link based on the transmission requirements. This approach is more advantageous than simply relying on a single link alone.

Then, we examine the effect of the distance from the BS to the DS on the increments of transmission success probability and transmission rate capacity, as depicted in Fig. \ref{biax-cre}. Notably, both the increments of transmission success probability and transmission rate capacity gradually increase, starting from zero and reaching their peak values. This observation demonstrates the importance of the space link in LEO satellite-assisted shore-to-ship communications, especially when the distance between the base station and the ship is long. 

\section{Conclusion}
In this paper, we have presented an analytical framework to obtain the end-to-end signal transmission performance for the LEO-SSCN that consists of an NSC, a BS, an ES, a DS, and LEO satellites modeled as a BPP uniformly distributed on a specific spherical surface. The NSC transmits signals to the DS through either the marine link or the space link subject to Rician fading and SR fading, respectively. For the threshold-based communication, the link selection relies on the relationship between the SNR of the pilot signal in the marine link and the predefined threshold value. The end-to-end transmission success probability and average transmission rate capacity are derived for an arbitrarily located DS on the sea surface by utilizing stochastic geometry, incorporating the Laplace transform of the interfering satellites' power. Furthermore, we introduce a tractable distance approximation approach to estimate the distance from the SS to the DS by using the distance from the ES to the SS in the constructed geometry structure among NSC, SS, and DS. Monte Carlo simulation results demonstrate the accuracy of our derivations. Based on the analytical framework, we investigate the effect of key parameters on the end-to-end signal transmission performance of the LEO-SSCN, including constellation size, constellation altitude, and the distance from the BS to the DS. We also demonstrate the performance improvement of the LEO-assisted transmission compared to the marine link transmission and examine the impact of various parameters on this improvement. The theoretical results match the simulation results, validating the accuracy of the derived analytical expressions. Notably, after incorporating the space link, the transmission success probability increases by 886${\rm{\% }}$ with a 13 dB predefined SNR (or SINR) threshold. This superior performance is attributed to the fact that the space link uses a wider bandwidth and greater power for signal transmission compared to the maritime link. It’s undeniable that the integration of the space link inevitably incurs additional expenses. This framework paves the way for more reliable integration of the LEO satellites and the existing maritime communication network.

Further research includes the study of a heterogeneous maritime network architecture with a diverse set of devices, such as unmanned aerial vehicles (UAVs), unmanned surface vehicles (USVs), and buoys, which characterize practical maritime scenarios. Further refinement of the analytical framework can be achieved by incorporating more practical path loss models, such as the two-ray model and three-ray model, and transforming the Cartesian coordinates into spherical coordinates in presenting the positions of LEO satellites, to provide a more realistic evaluation of transmission performance. To expand the analytic framework, it is necessary to consider multi-tier LEO satellite constellations with different altitudes, transmit powers, and interfering satellites from multiple layers. 

%Additionally, evaluating the impact of different communication schemes based on maritime communication demands would provide valuable insights into crucial performance metrics. 
% Further refinement of the analytical framework can be achieved by incorporating more practical path loss models, such as the two-ray model and three-ray model, to provide a more realistic evaluation of the transmission performance.

{\appendices
\section*{Appendix}
\subsection{Proof of Theorem 1}
\textit{Proof}: To obtain (\ref{coveragelink1}), we start with the definition of the success probability in the marine link
\begin{equation} \label{coveragelink1-proof}
\begin{aligned}
 {P_{{\rm{bd}}}}(\tau ) &=\mathbb{P} \left( {{\rm{SN}}{{\rm{R}}_{{\rm{bd}}}} > \tau } \right)\\
 &=\mathbb{P} \left( {\frac{{{p_{{\rm{bd}}}}{G_{{\rm{bd}}}}{{\left| {{H_{{\rm{bd}}}}} \right|}^2}R_{{\rm{bd}}}^{ - {\alpha _{{\rm{bd}}}}}}}{{\sigma _{{\rm{bd}}}^2}} > \tau } \right)\\
 &=\mathbb{P} \left( {\left| {{H_{{\rm{bd}}}}} \right| > \sqrt {\frac{{\tau \sigma _{{\rm{bd}}}^2R_{{\rm{bd}}}^{{\alpha _{{\rm{bd}}}}}}}{{{p_{{\rm{bd}}}}{G_{{\rm{bd}}}}}}} } \right)\\
& \underline{\underline {\left( a \right)}} {Q_1}\left( {\frac{\upsilon }{\rho },\frac{1}{\rho }\sqrt {\frac{{\tau \sigma _{{\rm{bd}}}^2R_{{\rm{bd}}}^{{\alpha _{{\rm{bd}}}}}}}{{{p_{{\rm{bd}}}}{G_{{\rm{bd}}}}}}} } \right),
\end{aligned}
\end{equation}
\noindent where $(a)$ is obtained by using the complementary CDF (CCDF) of the Rician fading model. % that can be expressed in (\ref{coveragelink1-proof}), 

Similarly, to obtain (\ref{coveragelink2}), we start with the definition of the success probability in the space link, which can be expressed in (\ref{coveragelink2-proof}). Note that ${\cal D} = \frac{{{{\left( {1 - m} \right)}_n}{{\left( { - \delta } \right)}^n}}}{{{{\left( {n!} \right)}^2}}}$, ${\cal J} = \frac{{n!}}{{l!}}{\left( {\beta  - \delta } \right)^{ - \left( {n + 1 - l} \right)}}$, $(a)$ is obtained by using the CCDF of the SR fading model, and $(b)$ from the PDF of ${R_{\rm{u}}}$ in Lemma \ref{Lemma-2}, and ${R_{\rm{d}}}$ is replaced by $\sqrt { {R_{\rm{u}}^2 + R_{{\rm{bd}}}^2} }$. In $(c)$, ${{\cal L}_{\left( {\sigma _{\rm{d}}^2 + I_s} \right)}}\left( {s_1\left( {\beta  - \delta } \right)} \right)$ represents the Laplace transform of the sum power of interfering satellites $I_s$ and received noise power $\sigma _{\rm{d}}^{\rm{2}}$ in the second time slot of the space link, which can be derived from the definition of Laplace transform
\begin{figure*}
\begin{equation}\label{coveragelink2-proof}
\begin{aligned}
&  {P_{{\rm{esd}}}}(\tau ) \\
&\buildrel \Delta \over = \mathbb{P}\left( {{\rm{SN}}{{\rm{R}}_{\rm{u}}} > \tau ,{\rm{ SIN}}{{\rm{R}}_{\rm{d}}} > \tau } \right)\\
&= \mathbb{E} {_{{{\left| {{H_{\rm{u}}}} \right|}^2},{R_{\rm{u}}}, {{\left| {{H_{\rm{d}}}} \right|}^2},I_s}}\left[ {\mathbb{P}\left( {\frac{{{p_{\rm{u}}}{G_{\rm{u}}}{{\left| {{H_{\rm{u}}}} \right|}^2}R_{\rm{u}}^{ - {\alpha}}}}{{\sigma _{\rm{u}}^{\rm{2}}}} > \tau ,{\rm{ }}\frac{{{p_{\rm{d}}}{G_{\rm{d}}}{{\left| {{H_{\rm{d}}}} \right|}^2}R_{\rm{d}}^{ - {\alpha}}\left( {{R_{\rm{u}}}} \right)}}{{\sigma _{\rm{d}}^2 + I_s}} > \tau } \right)} \right]\\
&= \mathbb{E} {_{{{\left| {{H_{\rm{u}}}} \right|}^2},{R_{\rm{u}}}, {{\left| {{H_{\rm{d}}}} \right|}^2},I_s}}\left[ {\mathbb{P}\left( {{{\left| {{H_{\rm{u}}}} \right|}^2} > \frac{{\tau \sigma _{\rm{u}}^{\rm{2}}R_{\rm{u}}^{{\alpha }}}}{{{p_{\rm{u}}}{G_{\rm{u}}}}}} \right) \times \mathbb{P}\left( {{{\left| {{H_{\rm{d}}}} \right|}^2} > \frac{{\tau R_{\rm{d}}^{{\alpha}}\left( {{R_{\rm{u}}}} \right)\left( {\sigma _{\mathop{\rm d}\nolimits} ^2{\rm{ + }}I_s} \right)}}{{{p_{\rm{d}}}{G_{\rm{d}}}}}} \right)} \right]\\
 &\underline{\underline {\left( a \right)}} \mathbb{E} {_{{R_{\rm{u}}},I_s}}\left[ {\left( {\mu \sum\limits_{n = 0}^{m - 1} {\cal D} \sum\limits_{l = 0}^n {{\cal J}} {{\left( {\frac{{\tau \sigma _{\rm{u}}^{\rm{2}}R_{\rm{u}}^{{\alpha}}}}{{{p_{\rm{u}}}{G_{\rm{u}}}}}} \right)}^l}{{\rm{e}}^{ - \left( {\beta  - \delta } \right)\frac{{\tau \sigma _{\rm{u}}^{\rm{2}}R_{\rm{u}}^{{\alpha}}}}{{{p_{\rm{u}}}{G_{\rm{u}}}}}}}} \right)} \right. \times  \left. \left( {\mu \sum\limits_{n = 0}^{m - 1} {\cal D} \sum\limits_{l = 0}^n {{\cal J}} {{\left( {\frac{{\tau R_{\rm{d}}^{{\alpha }}\left( {{R_{\rm{u}}}} \right)\left( {\sigma _{\mathop{\rm d}\nolimits} ^2{\rm{ + }}I_s} \right)}}{{{p_{\rm{d}}}{G_{\rm{d}}}}}} \right)}^l}{{\rm{e}}^{ - \left( {\beta  - \delta } \right)\frac{{\tau R_{\rm{d}}^{{\alpha}}\left( {{R_{\rm{u}}}} \right)\left( {\sigma _{\mathop{\rm d}\nolimits} ^2{\rm{ + }}I_s} \right)}}{{{p_{\rm{d}}}{G_{\rm{d}}}}}}}} \right) \right] \\
&\underline{\underline {\left( b \right)}} \mathbb{E}_{{I_s}}\Bigg[ \int_{{r_{\min }}}^{{r_{\max }}} 
\left( {\mu \sum\limits_{n = 0}^{m - 1} {\cal D} \sum\limits_{l = 0}^n {\cal J} {{\left( {\frac{{\tau \sigma _{\rm{u}}^{\rm{2}}R_{\rm{u}}^{{\alpha}}}}{{{p_{\rm{u}}}{G_{\rm{u}}}}}} \right)}^l}{{\rm{e}}^{ - \left( {\beta  - \delta } \right)\frac{{\tau \sigma _{\rm{u}}^{\rm{2}}R_{\rm{u}}^{{\alpha}}}}{{{p_{\rm{u}}}{G_{\rm{u}}}}}}}} \right) \\
&\quad \times \left( {\mu \sum\limits_{n = 0}^{m - 1} {\cal D} \sum\limits_{l = 0}^n {\cal J} {{\left( {\frac{{\tau {\left( {r_{\rm{u}}^2 + R_{{\rm{bd}}}^2} \right)^{\frac{{{\alpha }}}{2}}}\left( {\sigma _{\mathop{\rm d}\nolimits} ^2{\rm{ + }}I_s} \right)}}{{{p_{\rm{d}}}{G_{\rm{d}}}}}} \right)}^l}{{\rm{e}}^{ - \left( {\beta  - \delta } \right)\frac{{\tau {\left( {r_{\rm{u}}^2 + R_{{\rm{bd}}}^2} \right)^{\frac{{{\alpha}}}{2}}}\left( {\sigma _{\mathop{\rm d}\nolimits} ^2{\rm{ + }}I_s} \right)}}{{{p_{\rm{d}}}{G_{\rm{d}}}}}}}} \right) 
\times {f_{{R_{\rm{u}}}}}\left( {{r_{\rm{u}}}} \right)d{r_{\rm{u}}}
\Bigg]\\
&= \mathbb{E}_{{I_s}}\left[ {\int_{{r_{\min }}}^{{r_{\max }}} {\left( {\mu \sum\limits_{n = 0}^{m - 1} {\cal D} \sum\limits_{l = 0}^n {\cal J} {{\left( {\frac{{\tau \sigma _{\rm{u}}^{\rm{2}}R_{\rm{u}}^{{\alpha}}}}{{{p_{\rm{u}}}{G_{\rm{u}}}}}} \right)}^l}{{\rm{e}}^{ - \left( {\beta  - \delta } \right)\frac{{\tau \sigma _{\rm{u}}^{\rm{2}}R_{\rm{u}}^{{\alpha}}}}{{{p_{\rm{u}}}{G_{\rm{u}}}}}}}} \right)} \left. {{\times \left( {\mu \sum\limits_{n = 0}^{m - 1} {\cal D} \sum\limits_{l = 0}^n {\cal J} {{\left( {s_1\left( {\sigma _{\mathop{\rm d}\nolimits} ^2{\rm{ + }}I_s} \right)} \right)}^l}{{\rm{e}}^{ - s_1\left( {\beta  - \delta } \right)\left( {\sigma _{\mathop{\rm d}\nolimits} ^2{\rm{ + }}I_s} \right)}}} \right){f_{{R_{\rm{u}}}}}\left( {{r_{\rm{u}}}} \right)d{r_{\rm{u}}}} } \right]} \right\}\\
&\underline{\underline {\left( c \right)}}  \int_{{r_{\min }}}^{{r_{\max }}} {{\left( {1 - \frac{{r_{\rm{u}}^2 - r_{\min }^2}}{{4{r_{\rm{e}}}{r_a}}}} \right)}^{N - 1}}\frac{{{r_{\rm{u}}}N}}{{2{r_{\rm{e}}}{r_a}}} \times {\left( {\mu \sum\limits_{n = 0}^{m - 1} {\cal D} \sum\limits_{l = 0}^n {\cal J} {{\left( {\frac{{\tau \sigma _{\rm{u}}^{\rm{2}}R_{\rm{u}}^{{\alpha}}}}{{{p_{\rm{u}}}{G_{\rm{u}}}}}} \right)}^l}{{\rm{e}}^{ - \left( {\beta  - \delta } \right)\frac{{\tau \sigma _{\rm{u}}^{\rm{2}}R_{\rm{u}}^{{\alpha}}}}{{{p_{\rm{u}}}{G_{\rm{u}}}}}}}} \right)}\\
& \quad {\times \left( {\mu \sum\limits_{n = 0}^{m - 1} {\cal D} \sum\limits_{l = 0}^n {\frac{{n!}}{{l!}}}  {\left( {\beta  - \delta } \right)^{ - \left( {n + 1 } \right)}}{{\left( { - s_1} \right)}^l}\frac{{{\partial ^{(l)}}}}{{\partial {s_1^{\left( l \right)}}}}{{\cal L}_{\left( {\sigma _{\rm{d}}^2 + I_s} \right)}}\left( {s_1\left( {\beta  - \delta } \right)} \right)} \right) d{r_{\rm{u}}}}.
\end{aligned}
\end{equation} 
\hrulefill%black line
\end{figure*}
\begin{equation} \label{coverageLaISd}
\resizebox{0.98\hsize}{!}{$
\begin{aligned}
&{{\cal L}_{\left( {\sigma _{\rm{d}}^2 + I_s} \right)}}\left( s_1\left( {\beta  - \delta } \right)\right) \\
& \buildrel \Delta \over = \mathbb{E}{_{I_s}}\left[ {{e^{ - s_1\left( {\beta  - \delta } \right)\left( {\sigma _{\rm{d}}^2 + I_s} \right)}}} \right]\\
& = {\cal A} \times \mathbb{E}{_{I_s}}\left[ {{e^{ - s_1\left( {\beta  - \delta } \right)I_s}}} \right]\\
&= {\cal A} \times \mathbb{E}{_{{N_{\rm{I}}},{\left| {{H_j}} \right|^2},{R_j}}}\left[ {\exp \left( { - s_1\left( {\beta  - \delta } \right)\mathop \sum \limits_{j = 1}^{{N_{\rm{I}}}} {p_{\rm{i}}}{G_{\rm{i}}}{\left| {{H_j}} \right|^2}R_j^{ - {\alpha}}} \right)} \right]\\
&\underline {\underline {(a)} } {\cal A} \times \mathbb{E}{_{{N_{\rm{I}}},{R_j}}}\left\{ {\prod\limits_{j = 1}^{{N_{\rm{I}}}} {\mathbb{E}{_{{{\left| {{H_j}} \right|}^2}}}\left[ {\exp \left( { - {s_1}(\beta  - \delta ){p_{\rm{i}}}{G_{\rm{i}}}{{\left| {{H_j}} \right|}^2}R_j^{ - \alpha }} \right)} \right]} } \right\}\\
&\underline {\underline {(b)} } {\cal A} \times \mathbb{E}{_{{N_{\rm{I}}}}}\left\{ {\prod\limits_{j = 1}^{{N_{\rm{I}}}} {\int_{{r_{\rm{u}}}}^{{r_{\max }}} \mathbb{E}{{_{{{\left| {{H_j}} \right|}^2}}}} } } \right.\left[ {\exp \left( { - {s_1}(\beta  - \delta ){p_{\rm{i}}}{G_{\rm{i}}}{{\left| {{H_j}} \right|}^2}r_j^{ - \alpha }} \right)} \right.\\
& \quad \left. {\left. { \times {f_{{R_j}\mid {R_{\rm{u}}}}}\left( {{r_j}\mid {r_{\rm{u}}}} \right)d{r_j}} \right]} \right\}\\
& \underline {\underline {(c)} } {\cal A} \times \mathbb{E}{_{{N_{\rm{I}}}}}\left\{ {\mathop \prod \limits_{j = 1}^{{N_{\rm{l}}}} \left[ {\frac{2}{{{P_{\rm{I}}}\left( {4{r_{\rm{e}}}{r_{\rm{a}}} - r_{\rm{u}}^2 + r_{\min }^2} \right)}}} \right.} \right.\\
& \quad \left. {\left. { \times \int_{{r_{\rm{u}}}}^{{r_{\max }}} \mathbb{E}{{_{{{\left| {{H_j}} \right|}^2}}}} \left[ {\exp \left( { - {s_1}(\beta  - \delta ){p_{\rm{i}}}{G_{\rm{i}}}{{\left| {{H_j}} \right|}^2}r_j^{ - \alpha }} \right)} \right]{r_j}d{r_j}} \right]} \right\}\\
& \underline{\underline {(d)}} {\cal A} \times \mathop \sum \limits_{{n_{\rm{I}}} = 0}^{\frac{N}{K} - 1} \left[ {\left( { {\begin{array}{*{20}{c}} {\frac{N}{K} - 1}\\ {{n_{\rm{I}}}} \end{array}} } \right)P_{\rm{I}}^{{n_{\rm{I}}}}{{\left( {1 - {P_{\rm{I}}}} \right)}^{\frac{N}{K} - 1 - {n_{\rm{I}}}}}{{\left( {\frac{2}{{\left( {4{r_{\rm{e}}}{r_{\rm{a}}} - r_{\rm{u}}^2 + r_{\min }^2} \right){P_{\rm{I}}}}}} \right)}^{{n_{\rm{I}}}}}} \right.\\
&\quad \left. { \times {{\left( {\int_{{r_{\rm{u}}}}^{{r_{\max }}} {{{\cal L}_{{\left| {{H_j}} \right|^2}}}\left( {s_1\left( {\beta  - \delta } \right){p_{\rm{i}}}{G_{\rm{i}}}r_j^{ - {\alpha}}} \right)} {r_j}d{r_j}} \right)}^{{n_{\rm{I}}}}}} \right].\\
\end{aligned}$}
\end{equation}

\noindent Note that ${\cal A} = {e^{ - {s_1}\sigma _{\rm{d}}^2(\beta  - \delta )}}$, $(a)$ follows from the i.i.d. distribution of ${\left| {{H_j}} \right|^2}$ and its independence from ${N_{\rm{I}}}$ and ${R_j}$, $(b)$ is obtained using the PDF of ${R_j}$ in Lemma \ref{Lemma-3}, $(c)$ is the averaging over the binomial random variable ${N_{\rm{I}}}$ with the probability ${P_{\rm{I}}}$, which is given in Lemma \ref{Lemma-4}, and in $(d)$, ${{\cal L}_{\left| {{H_j}} \right|^2}}\left( {{s_1}{p_{\rm{i}}}{G_{\rm{i}}}r_j^{ - \alpha }} \right)$ is the Laplace transform of the channel gain from ISs to the DS. Since the SR fading model is a special case of the $\kappa-\mu$ fading model \cite{pc1}, we employ the Laplace transform of the $\kappa-\mu$ fading model to represent the Laplace transform of the SR fading model. It is equivalent to the SR fading model under the conditional that $\mu {\rm{ = }}1$, $\kappa  = {K_{\kappa  - \mu }}$ and $m$ takes the same value as the SR fading model \cite{pc1}. Then, by substituting the equivalent parameters to the Laplace function of the $\kappa-\mu$ fading model, we can obtain 
\begin{equation}
\resizebox{0.97\hsize}{!}{$
\begin{aligned}
{{\cal L}_{\left| {{H_j}} \right|^2}}\left( {{s_1}{p_{\rm{i}}}{G_{\rm{i}}}r_j^{ - \alpha }} \right)=&{{{\left( {1 + \frac{{\overline h s_1\left( {\beta  - \delta } \right){p_{\rm{i}}}{G_{\rm{i}}}r_j^{ - {\alpha}}}}{{1 + {{K_{\kappa  - \mu }}}}}} \right)}^{{m} - 1}}}\\
& \times {{\left( {1 + \frac{{\left( {{{K_{\kappa  - \mu }}} + {m}} \right)\overline h s_1\left( {\beta  - \delta } \right){p_{\rm{i}}}{G_{\rm{i}}}r_j^{ - {\alpha}}}}{{\left( {1 + {{K_{\kappa  - \mu }}}}  \right)} {m}}} \right)}^{ - {m}}},
\end{aligned}$}
\end{equation}
\noindent where $s_1 = \frac{{\tau {{\left( {r_{\rm{u}}^2 + R_{{\rm{bd}}}^2} \right)}^{\frac{{{\alpha}}}{2}}}}}{{{p_{\rm{d}}}{G_{\rm{d}}}}}$, $\overline h  = \mathbb{E} \left[ h \right]$ means the average value of the $\kappa  - \mu $ fading, ${m}$ indicates the Nakagami parameter, and ${K_{\kappa  - \mu }}$ represents the ratio between the average power of the LoS components $\Omega$ and the total power of the multi-path component $2b$, that is ${K_{\kappa  - \mu }} = {\Omega  \mathord{\left/
 {\vphantom {\Omega  {\left( {2b} \right)}}} \right.  \kern-\nulldelimiterspace} {\left( {2b} \right)}}$. \hfill $\square$

%In (\ref{coverageLaISd}),  which can be expressed in (\ref{cap-dir-proof}). Note that 

\subsection{Proof of Theorem 2}
\textit{Proof}: To obtain ({\ref{capacitydir}}), we start with the Shannon capacity in the marine link
\begin{align*}
&\int_{\sqrt {\frac{{\tau \sigma _{{\rm{bd}}}^2R_{{\rm{bd}}}^{{\alpha _{{\rm{bd}}}}}}}{{{p_{{\rm{bd}}}}{G_{{\rm{bd}}}}}}} }^\infty  {{f_{{H_{{\rm{bd}}}}}}\left( x \right){C_{{\rm{bd}}}}dx}  \\
& \buildrel \Delta \over = {B_{{\rm{bd}}}}\mathbb{E}{_{{H_{{\rm{bd}}}}}}\left[ {{{\log }_2}\left( {1 + \frac{{{p_{{\rm{bd}}}}{G_{{\rm{bd}}}}R_{{\rm{bd}}}^{ - {\alpha _{{\rm{bd}}}}}{{\left| {{H_{{\rm{bd}}}}} \right|}^2}}}{{\sigma _{{\rm{bd}}}^{\rm{2}}}}} \right)} \right]
\end{align*}
\begin{equation} \label{cap-dir-proof}
		\begin{aligned}
			& =  {B_{{\rm{bd}}}}{\int_{\sqrt {\frac{{\tau \sigma _{{\rm{bd}}}^2R_{{\rm{bd}}}^{{\alpha _{{\rm{bd}}}}}}}{{{p_{{\rm{bd}}}}{G_{{\rm{bd}}}}}}} }^\infty  {{{\log }_2}\left( {1 + \frac{{{p_{{\rm{bd}}}}{G_{{\rm{bd}}}}R_{{\rm{bd}}}^{ - {\alpha _{{\rm{bd}}}}}{x^2}}}{{\sigma _{{\rm{bd}}}^{\rm{2}}}}} \right)}{f_{{H_{{\rm{bd}}}}}}\left( x \right)dx}\\
			& \underline{\underline {\left( {a} \right)}} {B_{{\rm{bd}}}}{\int_{\sqrt {\frac{{\tau \sigma _{{\rm{bd}}}^2R_{{\rm{bd}}}^{{\alpha _{{\rm{bd}}}}}}}{{{p_{{\rm{bd}}}}{G_{{\rm{bd}}}}}}} }^\infty  {{{\log }_2}\left( {1 + \frac{{{p_{{\rm{bd}}}}{G_{{\rm{bd}}}}R_{{\rm{bd}}}^{ - {\alpha _{{\rm{bd}}}}}{x^2}}}{{\sigma _{{\rm{bd}}}^{\rm{2}}}}} \right)} }\\
			&\quad \times {\frac{x}{{{\rho ^2}}}\exp \left( {\frac{{ - \left( {{x^2} + {v ^2}} \right)}}{{2{\rho ^2}}}} \right){I_0}\left( {\frac{{xv }}{{{\rho ^2}}}} \right)dx},
		\end{aligned}
\end{equation}
\noindent where $(a)$ is obtained by using the PDF of the Rician fading model.

To obtain (${\ref{capacityre}}$), we start from the Shannon capacity for the smaller values of the ${\rm{SN}}{{\rm{R}}_{\rm{u}}}$ received at the SS and the ${\rm{SIN}}{{\rm{R}}_{\rm{d}}}$ received at the DS of the space link in (\ref{capacityre-derive}). Note that ${s_2} = \frac{{\left( {{2^t} - 1} \right){\left( {r_{\rm{u}}^{\rm{2}} + R_{{\rm{bd}}}^2} \right)^{\frac{\alpha }{2}}}}}{{{p_{\rm{d}}}{G_{\rm{d}}}}}$ and ${\cal D} = \frac{{{{\left( {1 - m} \right)}_n}{{\left( { - \delta } \right)}^n}}}{{{{\left( {n!} \right)}^2}}}$; $(a)$ is derived by applying the total probability formula, which implies that ${C_{{\rm{esd}}}}$ can be expanded as the sum of Shannon capacity subject to the condition of ${\rm{SN}}{{\rm{R}}_{\rm{u}}} > {\rm{SIN}}{{\rm{R}}_{\rm{d}}}$ and ${\rm{SN}}{{\rm{R}}_{\rm{u}}} < {\rm{SIN}}{{\rm{R}}_{\rm{d}}}$; for $(b)$, we approximate the total capacity of the space link by using the capacity of the second time slot. This is due to the fact that the ES equipped with a higher power transmitter has a sufficient energy supply, resulting in its transmit power usually being greater than that of the SS. In addition, interference generated by LEO satellites only exists in the second time slot, not in the first time slot, which greatly affects the SINR received at the DS. Extensive Monte Carlo simulations also indicate that the SNR received at the SS in the first time slot is always greater than the SINR received at the DS in the second time slot, referring to the practical network parameter settings. Therefore, using the capacity of the second time slot of the space link to approximate the total capacity of the space link is a reasonable and tractable approach; $(c)$ follows from the fact that for a positive random variable $X$, $\mathbb{E}\left[ X \right] = \int_{t > 0} \mathbb{P}{\left( {X > t} \right)dt}$; $(d)$ is obtained by substituting the PDF of the Rician fading model in (${\ref{pdfrician}}$) and the PDF of ${R_{\rm{u}}}$ in Lemma \ref{Lemma-2}. \hfill $\square$ 
%, which is shown in Eq. (${\ref{capacityre-derive}}$), , shown at the top of the next page
\begin{figure*}
\begin{equation}\label{capacityre-derive}
	\begin{aligned}
		& \int_0^{\sqrt{\frac{\tau \sigma_{\rm bd}^2 R_{\rm bd}^{\alpha_{\rm bd}}}{p_{\rm bd} G_{\rm bd}}}} 
		f_{H_{\rm bd}}(x)\, C_{\rm esd}\, dx \\
		& \overset{\Delta}{=} 
		\int_0^{\sqrt{\frac{\tau \sigma_{\rm bd}^2 R_{\rm bd}^{\alpha_{\rm bd}}}{p_{\rm bd} G_{\rm bd}}}} 
		f_{H_{\rm bd}}(x)\, B_{\rm esd}\,
		\mathbb{E}_{|H_{\rm u}|^2, R_{\rm u}, |H_{\rm d}|^2, I_{\rm s}}
		\left\{
		\log_2 \left[
		1 + \min \left(
		\frac{p_{\rm u} G_{\rm u} |H_{\rm u}|^2 R_{\rm u}^{-\alpha}}{\sigma_{\rm u}^2},
		\frac{p_{\rm d} G_{\rm d} |H_{\rm d}|^2 R_{\rm d}^{-\alpha}(R_{\rm u})}{\sigma_{\rm d}^2 + I_{\rm s}}
		\right)
		\right]
		\right\} dx \\
		& \underline{\underline{(a)}}\ 
		\int_0^{\sqrt{\frac{\tau \sigma_{\rm bd}^2 R_{\rm bd}^{\alpha_{\rm bd}}}{p_{\rm bd} G_{\rm bd}}}} 
		f_{H_{\rm bd}}(x)
		\Bigg\{
		B_{\rm esd}\,
		\mathbb{P} \left(
		\frac{p_{\rm u} G_{\rm u} |H_{\rm u}|^2 R_{\rm u}^{-\alpha}}{\sigma_{\rm u}^2} 
		< 
		\frac{p_{\rm d} G_{\rm d} |H_{\rm d}|^2 R_{\rm d}^{-\alpha}(R_{\rm u})}{\sigma_{\rm d}^2 + I_{\rm s}}
		\right)
		\mathbb{E}_{|H_{\rm u}|^2, R_{\rm u}} 
		\left[
		\log_2 \left(
		1 + \frac{p_{\rm u} G_{\rm u} |H_{\rm u}|^2 R_{\rm u}^{-\alpha}}{\sigma_{\rm u}^2}
		\right)
		\right] \\
		& \quad +
		B_{\rm esd}\,
		\mathbb{P} \left(
		\frac{p_{\rm u} G_{\rm u} |H_{\rm u}|^2 R_{\rm u}^{-\alpha}}{\sigma_{\rm u}^2}
		>
		\frac{p_{\rm d} G_{\rm d} |H_{\rm d}|^2 R_{\rm d}^{-\alpha}(R_{\rm u})}{\sigma_{\rm d}^2 + I_{\rm s}}
		\right)
		\mathbb{E}_{|H_{\rm d}|^2, R_{\rm u}, I_{\rm s}}
		\left[
		\log_2 \left(
		1 + \frac{p_{\rm d} G_{\rm d} |H_{\rm d}|^2 R_{\rm d}^{-\alpha}(R_{\rm u})}{\sigma_{\rm d}^2 + I_{\rm s}}
		\right)
		\right]
		\Bigg\} dx \\
		& \overset{(b)}{\approx}
		\int_0^{\sqrt{\frac{\tau \sigma_{\rm bd}^2 R_{\rm bd}^{\alpha_{\rm bd}}}{p_{\rm bd} G_{\rm bd}}}} 
		f_{H_{\rm bd}}(x)\, B_{\rm esd}\,
		\mathbb{E}_{|H_{\rm d}|^2, R_{\rm u}, I_{\rm s}}
		\left[
		\log_2 \left(
		1 + \frac{p_{\rm d} G_{\rm d} |H_{\rm d}|^2 R_{\rm d}^{-\alpha}(R_{\rm u})}{\sigma_{\rm d}^2 + I_{\rm s}}
		\right)
		\right] dx \\
		& \underline{\underline{(c)}}\ 
		\int_0^{\sqrt{\frac{\tau \sigma_{\rm bd}^2 R_{\rm bd}^{\alpha_{\rm bd}}}{p_{\rm bd} G_{\rm bd}}}} 
		f_{H_{\rm bd}}(x)\, B_{\rm esd}
		\int_{r_{\min}}^{r_{\max}}\!\!
		\int_0^\infty 
		\mathbb{E}_{|H_{\rm d}|^2, I_{\rm s}}
		\left[
		\mathbb{P} \left(
		|H_{\rm d}|^2 > 
		\frac{(2^t - 1)(\sigma_{\rm d}^2 + I_{\rm s}) R_{\rm d}^{\alpha}(R_{\rm u})}{p_{\rm d} G_{\rm d}}
		\right)
		\right] dt\,
		f_{R_{\rm u}}(r_{\rm u})\, dr_{\rm u}\, dx \\
		& = \int_0^{\sqrt{\frac{\tau \sigma_{\rm bd}^2 R_{\rm bd}^{\alpha_{\rm bd}}}{p_{\rm bd} G_{\rm bd}}}} 
		f_{H_{\rm bd}}(x)\, B_{\rm esd}
		\int_{r_{\min}}^{r_{\max}}
		\int_0^\infty 
		\mathbb{E}_{I_{\rm s}}
		\left[
		\mu \sum_{n=0}^{m-1} \mathcal{D} 
		\sum_{l=0}^n \mathcal{J}
		\left[
		s_2 (\sigma_{\rm d}^2 + I_{\rm s})
		\right]^l
		e^{-(\beta - \delta) s_2 (\sigma_{\rm d}^2 + I_{\rm s})}
		\right] dt\,
		f_{R_{\rm u}}(r_{\rm u})\, dr_{\rm u}\, dx \\
		& \underline{\underline{(d)}}\ 
		B_{\rm esd}
		\int_0^{\sqrt{\frac{\tau \sigma_{\rm bd}^2 R_{\rm bd}^{\alpha_{\rm bd}}}{p_{\rm bd} G_{\rm bd}}}}
		\int_{r_{\min}}^{r_{\max}}
		\int_0^\infty
		\mu \sum_{n=0}^{m-1} \mathcal{D}
		\sum_{l=0}^n 
		\frac{n!}{l!} (\beta - \delta)^{-(n+1)}
		(-s_2)^l
		\frac{\partial^{(l)}}{\partial s_2^{(l)}}
		\mathcal{L}_{\sigma_{\rm d}^2 + I_{\rm s}} \left( s_2 (\beta - \delta) \right)
		dt \\
		& \quad \times 
		\left(1 - \frac{r_{\rm u}^2 - r_{\min}^2}{4 r_{\rm e} r_a} \right)^{N-1}
		\frac{N r_{\rm u}}{2 r_{\rm e} r_a}
		f_{H_{\rm bd}}(x)\,
		\frac{x}{\rho^2}
		\exp \left( -\frac{x^2 + v^2}{2\rho^2} \right)
		I_0\left( \frac{x v}{\rho^2} \right)
		dr_{\rm u}\, dx .
	\end{aligned}
\end{equation}

\hrulefill%black line
\end{figure*}

%\noindent %, ${\cal G} = {\left( {\beta  - \delta } \right)^{ - \left( {n + 1 - l} \right)}}$ \left\{ {} \right\}
This concludes the proof.}

%Due to the dense deployment of the LEO satellites, the SS can be approximately regarded as directly above the NSC. On this basis, the distance from the ES to the SS and the distance from the SS to the DS are much larger than that from the BS to the DS, i.e., ${R_{\rm{u}}} \gg {R_{{\rm{bd}}}}$ and ${R_{\rm{d}}} \gg {R_{{\rm{bd}}}}$, thus the triangle among NSC, SS and DS in the constructed geometric structure can be approximately regarded as a right triangle. Therefore, we can utilize the Pythagorean theorem to approximate ${R_{\rm{d}}}$ by ${R_{\rm{u}}}$, that is $R_{\rm{d}}^{{\alpha}} \approx {\left( {R_{{\rm{u}}}^2 + R_{{\rm{bd}}}^2} \right)^{\frac{{{\alpha}}}{2}}}$. In (c), ${{\cal L}_{\left( {\sigma _{\rm{d}}^2 + I_s} \right)}}\left( {s_1\left( {\beta  - \delta } \right)} \right)$ represents the Laplace transform of the sum power of interfering satellites.

%\section*{Acknowledgments}
%The work was supported by the National Key Research and Development Program of China (No: 2019YFE0111600), National Natural Science Foundation of China (No: 61971083 and No: 51939001) and LiaoNing Revitalization Talents Program (No: XLYC2002078).

%\begin{thebibliography}{l}
\bibliographystyle{IEEEtran}
\bibliography{ref2025.bib} 

%\end{thebibliography}

%\newpage

%Different transmit power levels are assigned to SS and IS signals, denoted as ${p_\rm{s}}$ and ${p_\rm{i}}$, such that ${p_\rm{s}}\ge{p_\rm{i}}$.
%\begin{figure}[h] 
 %\center{\includegraphics[scale=0.6]{Fig.2-v6.pdf}} 
 %\caption{\label{1} The geometric structure of the LEO-SSCN.} 
 	%\label{fig:2}
%\end{figure}

\begin{comment}

\section{Biography Section}
If you have an EPS/PDF photo (graphicx package needed), extra braces are
 needed around the contents of the optional argument to biography to prevent
 the LaTeX parser from getting confused when it sees the complicated
 $\backslash${\tt{includegraphics}} command within an optional argument. (You can create
 your own custom macro containing the $\backslash${\tt{includegraphics}} command to make things
 simpler here.)
 
\vspace{11pt}
\end{comment}

%\bf{If you include a photo:}\vspace{-33pt}
\begin{IEEEbiography}
[{\includegraphics[width=1in,height=1.25in,clip,keepaspectratio]{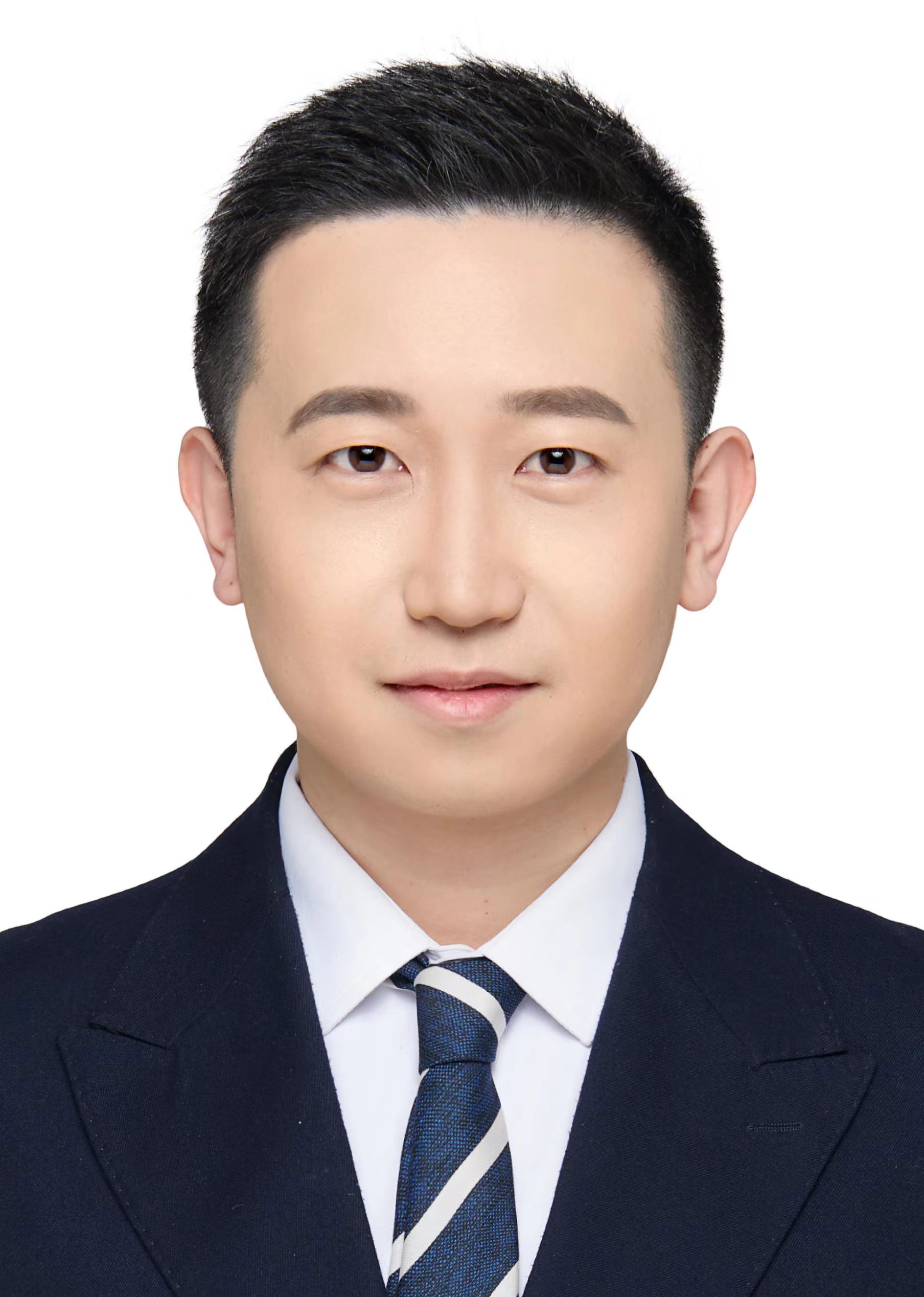}}]{Xu Hu}
[S’20] received the B.S. degree from Dalian Polytechnic University, Dalian, China, in 2017, and received the M.S. degree from Dalian Maritime University, Dalian, China, in 2020. He is currently pursuing the Ph.D. degree in Information Science and Technology College at Dalian Maritime University, Dalian, China. His research interests include maritime communication networks, space-air-ground-sea integrated communication networks, and stochastic geometry.
\end{IEEEbiography}

\begin{IEEEbiography}
[{\includegraphics[width=1in,height=1.25in,clip,keepaspectratio]{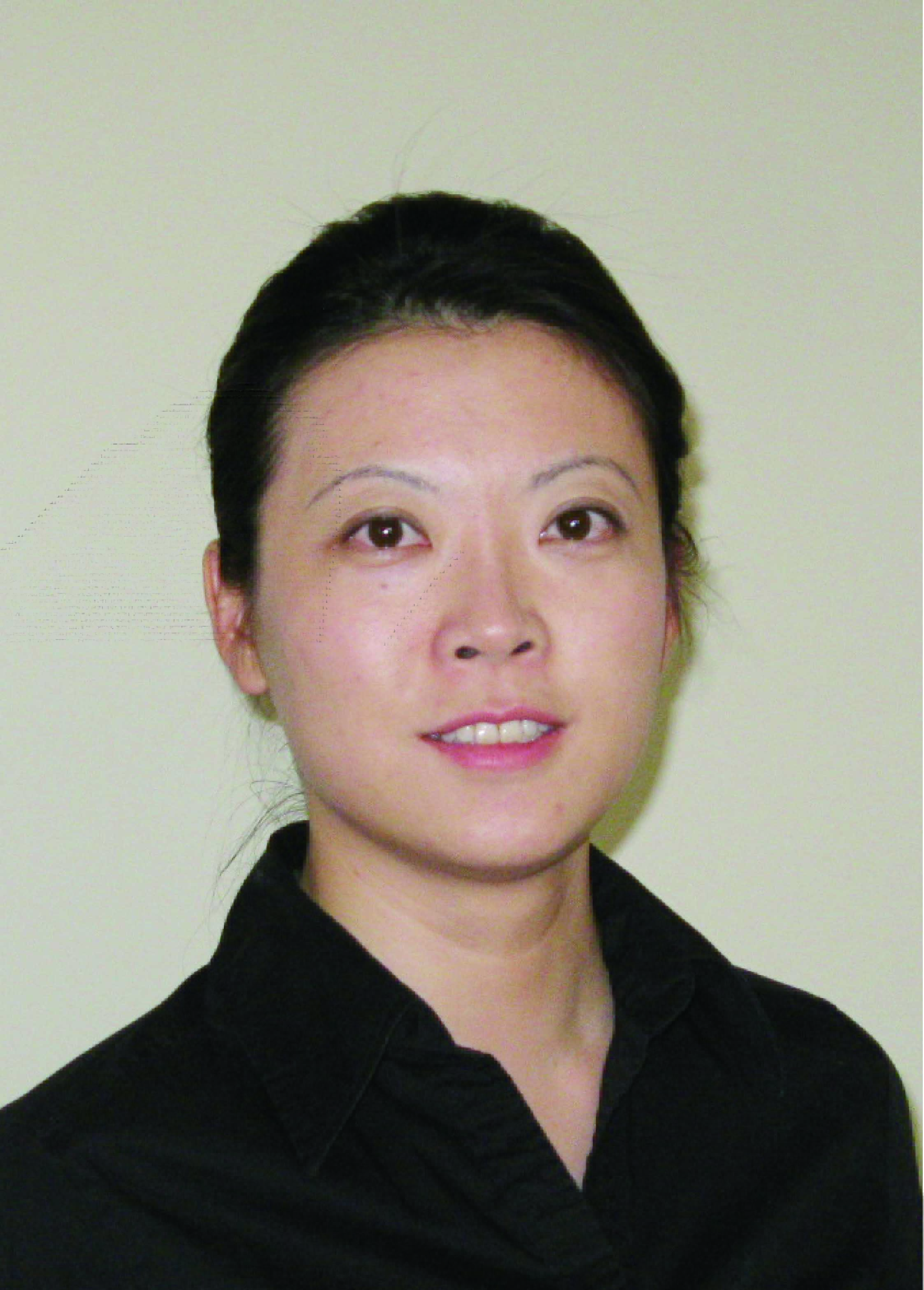}}]{Bin Lin}
[M’12-SM’21] received the B.S. and M.S. degrees from Dalian Maritime University, Dalian, China, in 1999 and 2003, respectively and the Ph.D. degree from the Broadband Communications Research Group, Department of Electrical and Computer Engineering, University of Waterloo, Waterloo, ON, Canada, in 2009. She is currently a Full Professor and the Dean of Communication Engineering with the College of Information Science and Technology, Dalian Maritime University. She has been a Visiting Scholar with George Washington University, Washington, DC, USA, from 2015 to 2016. Her current research interests include wireless communications, network dimensioning and optimization, resource allocation, artificial intelligence, maritime communication networks, and Internet of Things. Prof. Lin is an associate editor of IEEE Transactions on Vehicular Technology and IEEE Internet of Things Journal.
\end{IEEEbiography}

\begin{IEEEbiography}
[{\includegraphics[width=1in,height=1.25in,clip,keepaspectratio]{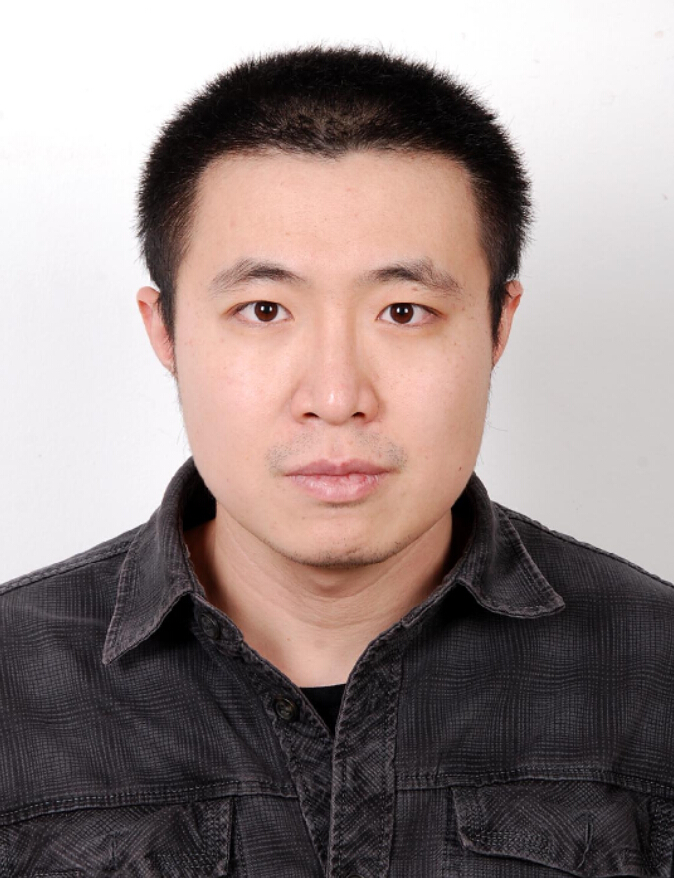}}]{Xiao Lu}
[M’20] received the B.Eng. degree in communication engineering from the Beijing University of Posts and Telecommunications, Beijing, China, the M.Eng. degree in computer engineering from Nanyang Technological University, Singapore, and the Ph.D. degree from the University of Alberta, Edmonton, AB, Canada. He is currently with Ericsson, Canada and York University, Canada. His current research interests include the design, analysis, and optimization of future generation cellular wireless networks. He was a recipient of the Best Paper Award from the IEEE Communication Society: Green Communications ${\rm{\& }}$ Computing Technical Committee in 2018 and the IEEE TCGCC Best Conference Paper.
\end{IEEEbiography}

\begin{IEEEbiography}
[{\includegraphics[width=1in,height=1.25in,clip,keepaspectratio]{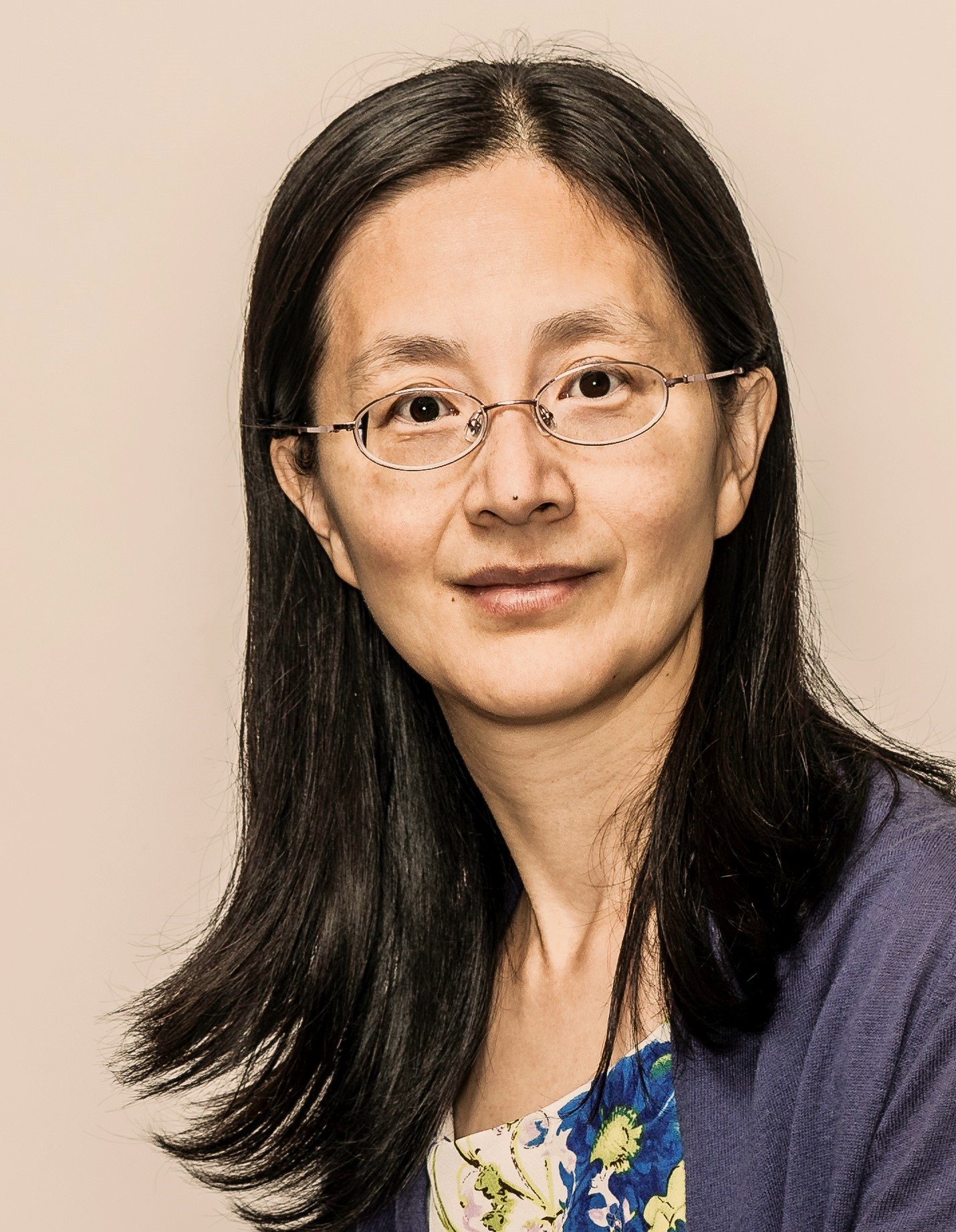}}]{Ping Wang}
[M’08-SM’14-F’22] is a Professor at the Department of Electrical Engineering and Computer Science, York University, and a Tier 2 York Research Chair. Prior to that, she was with Nanyang Technological University, Singapore, from 2008 to 2018.  Her research interests are mainly in radio resource allocation, network design, performance analysis and optimization for wireless communication networks, mobile cloud computing and the Internet of Things. Her recent works focus on integrating Artificial Intelligence (AI) techniques into communications networks. Her scholarly works have been widely disseminated through top-ranked IEEE journals/conferences, received 28,000+ citations, and received the Best Paper Awards from IEEE prestigious conference WCNC in 2012, 2020 and 2022, from IEEE Communication Society: Green Communications ${\rm{\& }}$ Computing Technical Committee in 2018, from IEEE flagship conference ICC in 2007. She has been serving as associate editor-in-chief for IEEE Communications Surveys ${\rm{\& }}$ Tutorials and an editor for several reputed journals, including IEEE Transactions on Wireless Communications. She is an IEEE Fellow and a \emph{Distinguished Lecturer of the IEEE Vehicular Technology Society}.
\end{IEEEbiography}

\begin{IEEEbiography}
[{\includegraphics[width=1in,height=1.25in,clip,keepaspectratio]{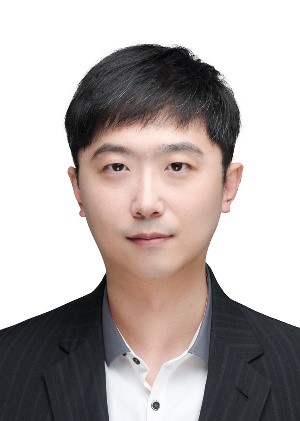}}]{Nan Cheng}
[M'16-SM’23] received the Ph.D. degree in Electrical and Computer Engineering from the University of Waterloo in 2016, and B.E. degree and the M.S. degree in Information and Telecommunications Engineering from Tongji University in 2009 and 2012, respectively. He worked as a Post-doctoral fellow with the Department of Electrical and Computer Engineering, University of Toronto, from 2017 to 2019. He is currently a professor with State Key Lab. of ISN and with School of Telecommunications Engineering, Xidian University, Shaanxi, China. He has published over 90 journal papers in IEEE Transactions and other top journals. He serves as associate editors for IEEE Internet of Things Journal, IEEE Transactions on Vehicular Technology, IEEE Open Journal of the Communications Society, and Peer-to-Peer Networking and Applications, and serves/served as guest editors for several journals. His current research focuses on B5G/6G, AI-driven future networks, and space-air-ground integrated network.
\end{IEEEbiography}

\begin{IEEEbiography}
[{\includegraphics[width=1in,height=1.25in,clip,keepaspectratio]{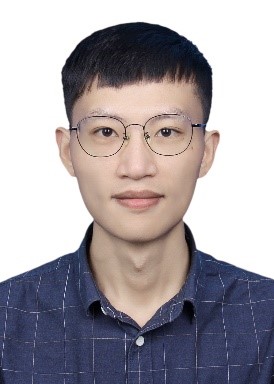}}]{Zhisheng Yin}
[M'20] received his Ph.D. degree from the School of Electronics and Information Engineering, Harbin Institute of Technology, Harbin, China, in 2020, and the B.E. degree from the Wuhan Institute of Technology, the B.B.A. degree from the Zhongnan University of Economics and Law, Wuhan, China, in 2012, and the M.Sc. degree from the Civil Aviation University of China, Tianjin, China, in 2016. From Sept. 2018 to Sept. 2019, Dr. Yin visited in BBCR Group, Department of Electrical and Computer Engineering, University of Waterloo, Canada. He is currently an Assistant Professor with School of Cyber Engineering, Xidian University, Xi'an, China. He is also an Associate Editor of IEEE Internet of Things Journal. His research interests include space-air-ground integrated networks, wireless communications, digital twin, and physical layer security.
\end{IEEEbiography}

\begin{IEEEbiography}
[{\includegraphics[width=1in,height=1.25in,clip,keepaspectratio]{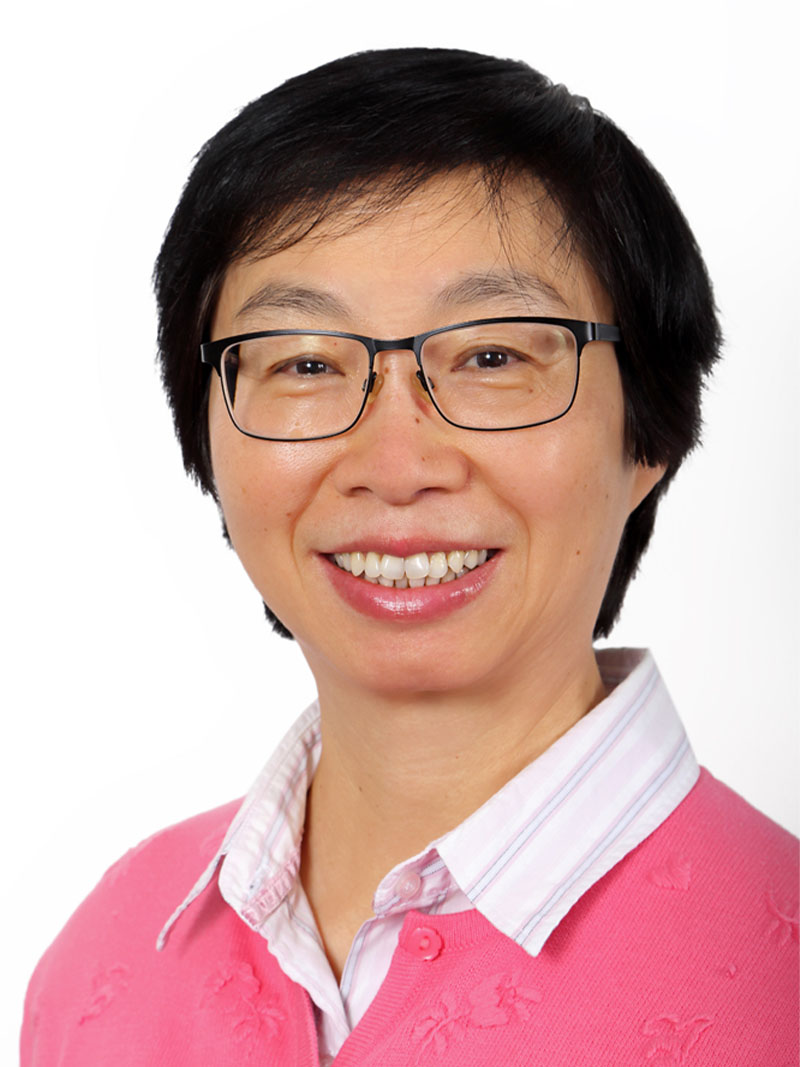}}]{Weihua Zhuang} 
[M’93-SM’01-F’08] received the B.Sc. and M.Sc. degrees from Dalian Maritime University, China, and the Ph.D. degree from the University of New Brunswick, Canada, all in electrical engineering. Since 1993, she has been a faculty member in the Department of Electrical and Computer Engineering, University of Waterloo, Canada, where she is a University Professor and a Tier I Canada Research Chair in wireless communication networks. Her current research focuses on network architecture, algorithms and protocols, and service provisioning in future communication systems. She is the recipient of Women’s Distinguished Career Award in 2021 from IEEE Vehicular Technology Society, R.A. Fessenden Award in 2021 from IEEE Canada, Award of Merit in 2021 from the Federation of Chinese Canadian Professionals (Ontario), and Technical Recognition Award in Ad Hoc and Sensor Networks in 2017 from IEEE Communications Society. Dr. Zhuang is a Fellow of the IEEE, Royal Society of Canada (RSC), Canadian Academy of Engineering (CAE), and Engineering Institute of Canada (EIC). She is the President and an elected member of the Board of Governors (BoG) of the IEEE Vehicular Technology Society. She was the Editor-in-Chief of IEEE Transactions on Vehicular Technology (2007-2013), an editor of IEEE Transactions on Wireless Communications (2005-2009), General Co-Chair of IEEE/CIC International Conference on Communications in China (ICCC) 2021, Technical Program Committee (TPC) Chair/Co-Chair of IEEE Vehicular Technology Conference 2017 Fall and 2016 Fall, TPC Symposia Chair of the IEEE Globecom 2011, and an IEEE Communications Society Distinguished Lecturer (2008-2011).
\end{IEEEbiography}

\begin{comment}
\begin{IEEEbiography}
Use $\backslash${\tt{begin\{IEEEbiography\}}} and then for the 1st argument use $\backslash${\tt{includegraphics}} to declare and link the author photo.
Hongda Wu is currently pursuing the Ph.D. degree, under the supervision of Prof. Ping Wang, at the Department of Electrical Engineering and Computer Science (Lassonde School of Engineering), York University, ON, Canada.
Use the author name as the 3rd argument followed by the biography text.
\end{IEEEbiography}

\vspace{11pt}

\bf{If you will not include a photo:}\vspace{-33pt}
\begin{IEEEbiographynophoto}{John Doe}
Use $\backslash${\tt{begin\{IEEEbiographynophoto\}}} and the author name as the argument followed by the biography text.
\end{IEEEbiographynophoto}
\end{comment}

\vfill

\end{document}